\documentclass[sigconf]{acmart}
\AtBeginDocument{%
  }

\copyrightyear{2025}
\acmYear{2025}
\setcopyright{acmlicensed}\acmConference[CHI '25]{CHI Conference on Human Factors in Computing Systems}{April 26-May 1, 2025}{Yokohama, Japan}
\acmBooktitle{CHI Conference on Human Factors in Computing Systems (CHI '25), April 26-May 1, 2025, Yokohama, Japan}
\acmDOI{10.1145/3706598.3713418}
\acmISBN{979-8-4007-1394-1/25/04}

\acmSubmissionID{9111}

\usepackage{graphicx}
\usepackage{stfloats}
\usepackage{xspace}
\usepackage{multirow}

\begin{document}

\title{\emph{GenColor}: Generative Color-Concept Association in Visual Design}

\author{Yihan Hou}
\email{yhou073@connect.hkust-gz.edu.cn}
\affiliation{%
  \institution{The Hong Kong University of Science and Technology (Guangzhou)}
  \country{Guangzhou, China}
}

\author{Xingchen Zeng}
\email{xzeng159@connect.hkust-gz.edu.cn}
\affiliation{
  \institution{The Hong Kong University of Science and Technology (Guangzhou)}
  \country{Guangzhou, China}
}

\author{Yusong Wang}
\email{ywang677@connect.hkust-gz.edu.cn}
\affiliation{%
  \institution{The Hong Kong University of Science and Technology (Guangzhou)}
  \country{Guangzhou, China}
}

\author{Manling Yang}
\email{myang838@connect.hkust-gz.edu.cn}
\affiliation{%
  \institution{The Hong Kong University of Science and Technology (Guangzhou)}
  \country{Guangzhou, China}
}

\author{Xiaojiao Chen}
\email{chenxiaojiao@zju.edu.cn}
\affiliation{%
  \institution{Zhejiang University}
  \country{Hangzhou, China}
}

\author{Wei Zeng}
\authornote{Wei Zeng is the corresponding author.}
\affiliation{%
  \institution{The Hong Kong University of Science and Technology (Guangzhou)}
  \country{Guangzhou, China}
}
\affiliation{%
  \institution{The Hong Kong University of Science and Technology}
  \country{Hong Kong SAR, China}
}


\definecolor{myred}{RGB}{243, 166, 148}
\definecolor{myblue}{RGB}{73, 148, 196}

\newcommand{\strike}[1]{\textcolor{gray}{\sout{#1}}}
\newcommand{\add}[1]{\textcolor{myblue}{#1}}
\newcommand{\re}[1]{\textcolor{black}{#1}}
\newcommand{\rec}[1]{\textcolor{black}{#1}}
\newcommand{\replace}[2]{\strikeg{#1 }\add{#2}}
\newcommand{\cmt}[2]{\textcolor{myred}{#1 }\add{#2}}
\newcommand{\Edit}[1]{\textcolor{orange}{Edit: #1}}

\newcommand{\q}[1]{\textit{``#1''}}
\newcommand{\qn}[1]{``#1''}
\newcommand{\eg}{\emph{e.g.}}
\newcommand{\ie}{\emph{i.e.}}
\newcommand{\vs}{\emph{vs.}}
\newcommand{\etal}{\emph{et al. }}

\newcommand{\strikeg}[1]{\textcolor{black}{\sout{#1}}}

\begin{abstract}
    Existing approaches for color-concept association typically rely on query-based image referencing, and color extraction from image references.
    However, these approaches are effective only for common concepts, and are vulnerable to unstable image referencing and varying image conditions.
    Our formative study with designers underscores the need for primary-accent color compositions and context-dependent colors (\eg, `clear' \vs `polluted' sky) in design.
    In response, we introduce a generative approach for mining semantically resonant colors leveraging images generated by text-to-image models.
    Our insight is that contemporary text-to-image models can resemble visual patterns from large-scale real-world data.
    The framework comprises three stages: \textit{concept instancing} produces generative samples using diffusion models, \textit{text-guided image segmentation} identifies concept-relevant regions within the image, and \textit{color association} extracts primary accompanied by accent colors.
    Quantitative comparisons with expert designs validate our approach's effectiveness, and we demonstrate the applicability through cases in various design scenarios and a gallery.
\end{abstract}




\maketitle

\section{Introduction}

\begin{figure*}[t]
    \centering
    \includegraphics[width=0.985\linewidth]{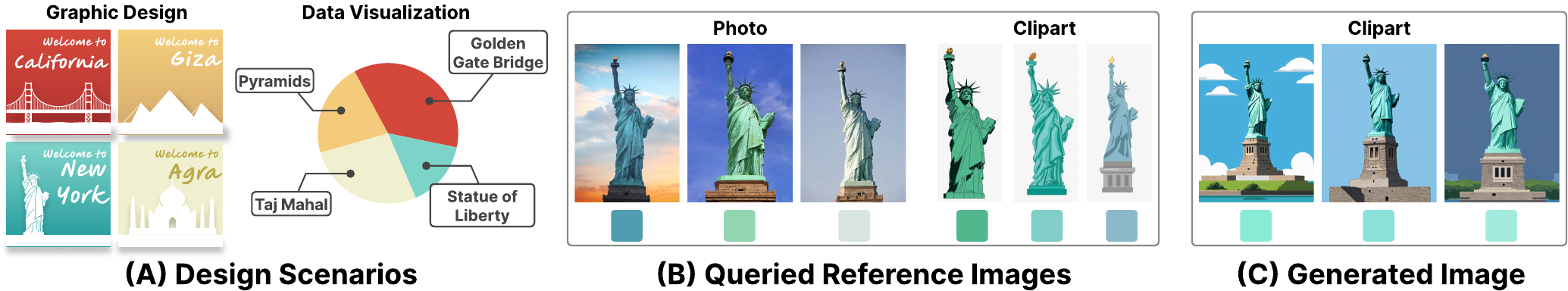}
    \caption{\re{Design scenarios where appropriate color-concept associations are effective to enhance visual communication: (A) graphic design and data visualization. These scenarios require color-concept association, which is preferable in generated images.} When comparing the queried and generated images for the \q{Statue of Liberty}, (B) colors extracted from the queried images (both photos and clipart) show significant variations, whereas (C) the generated images provide a more consistent color representation.}
    \Description{Design scenarios where appropriate color-concept associations are effective to enhance visual communication: (A) graphic design and data visualization. These scenarios require color-concept association, which is preferable in generated images. When comparing the queried and generated images for the \q{Statue of Liberty}, (B) colors extracted from the queried images (both photos and clipart) show significant variations, whereas (C) the generated images provide a more consistent color representation.}
    \label{fig:case_in_intro}
\end{figure*}

Color is a fundamental element in visual design, serving as a powerful approach to convey meaning and represent specific concepts~\cite{jahanian2017colors, chen2021investigation}.
\re{Many colors have well-established associations with particular semantics, enabling designers to enhance visual communication~\cite{wierzicka1990meaning}. 
These associations are valuable in domains such as data visualization~\cite{lin2013selecting,setlur2016linguistic,hu2023self} and graphic design~\cite{jahanian2017colors,shi2023stijl,hegemann2024palette}, where color associations are essential to represent categories or themes.
For instance, as illustrated in Figure~\ref{fig:case_in_intro} (A), a set of tourist pins for different cities can use colors linked to landmarks, such as red for the Golden Gate Bridge or sandy yellow for the Pyramids. 
Similarly, in data visualization, these concept-associated colors help categorize information effectively, enhancing clarity and engagement.}
However, selecting the satisfied color can be laborious, especially when designers are dealing with unfamiliar concepts or striving to align colors with specific emotions or contextual conditions.
For instance, when selecting colors for a group of historical landmarks, designers might generally infer the desired tone for the \q{Statue of Liberty} based on common sense (\eg, green or blue for the widely recognized appearance).
Careful selection is crucial to ensure that the chosen colors not only reflect the intended concept (such as the \q{Statue of Liberty}) but also evoke the desired context of emotions (\eg, \q{bringing light and hope}).
\re{The process remains challenging, particularly for novice designers.}

\re{This motivated the development of various methods to facilitate color-concept associations by providing external evidence and references to support designers' color design decisions~\cite{chen2021investigation,hegemann2024palette}.}
Crowdsourcing efforts~\cite{xkcd2010color} leverage collective intelligence to determine color associations; however, these methods can be time-consuming and resource-intensive.
In response, computational approaches quantifying color-concept associations have emerged.
These methods typically involve either mining color concept co-occurrences from large text corpora or identifying colors from image references retrieved from image databases~\cite{setlur2016linguistic,lin2013selecting,rathore2020estimating,hu2023self}.
Notably, this approach mirrors conventional practices used by designers who often search for color-referencing images from online resources.

However, these methods encounter significant limitations from various perspectives.
First, the linguistic methods in the initial stage are limited to common concepts and may find it challenging to capture context-dependent associations, as N-gram may be sparse for less frequent concepts~\cite{rathore2020estimating}.
For instance, while the concept of the \q{Statue of Liberty} is popular, the specific phrase \q{Statue of Liberty with the feeling of light and hope} could be rare.
Secondly, obtaining representative images with consistent color distributions from image databases for a given concept can be challenging.
Queried images can display significant visual discrepancies due to various external factors such as lighting conditions, image quality, and stylistic variations.
For instance, photos of the statue retrieved from Google Image Search exhibit notable differences, as illustrated in Figure~\ref{fig:case_in_intro}.
Some studies~\cite{setlur2016linguistic, lin2013selecting} attempt to refine the query results by incorporating terms like \q{clipart}, yet the outcomes still show considerable variations.
These complex color disparities in the queried results make it challenging to determine the most representative color for the \q{Statue of Liberty}.
A flexible and robust computational approach is needed for mining color-concept associations.

We present a novel framework, \textit{GenColor}, which utilizes a generative approach to extract semantically meaningful colors from images produced by text-to-image (T2I) models.
\re{By providing designers with a flexible and supportive framework, GenColor enables them to explore color concepts and select contextually appropriate color palettes, serving as a reference for their design decisions.
Our work is motivated by the insight that contemporary T2I models are trained on large-scale real-world data, and the generated high-quality images can resemble visual patterns in the real world~\cite{orgad2023editing,li2024sd4match,chen2024oppurtunities}.}
As such, T2I models can serve as a data mining tool to uncover real-world visual patterns~\cite{ioannis2024diffusion}.
The design of \textit{GenColor} is informed by design goals from a preliminary study on designers' workflows and their expectations for \re{a supportive tool} for color-concept association (Sect.~\ref{sec:preliminary}).
The \textit{GenColor} framework is implemented in three stages:
1) \textit{Conceptual Instancing} (Sect.~\ref{ssec:conceptual-instancing}) flexibly generates representative image samples of the corresponding context-dependent concept through the stable diffusion model,
2) \textit{Text-guided Image Segmentation} (Sect.~\ref{ssec:image-segmentation}) identifies semantic-related regions within the image, and
3) \textit{Color Association} (Sect.~\ref{ssec:color-association}) extracts primary-accent color compositions.

To evaluate the effectiveness of \textit{GenColor}, we collect a designer baseline dataset for concept coloring of 36 common concepts (Sect.~\ref{ssec:designer-baseline}), and conduct a quantitative comparison with existing query-based approaches (Sect.~\ref{ssec:quantitative-evaluation}).
The results show that \textit{GenColor} produces color compositions that closely align with the designers' choices and outperform query-based approaches in terms of representativeness and designer preference.
\re{This indicates that the generative framework offers two primary advantages over retrieval-based approaches:
1) \emph{feasibility and flexibility} in generating representative images for context-dependent concepts; and 2) \emph{robustness and effectiveness} in managing variations in the generated images, such as controlling lighting conditions and style.}
We also demonstrate the application scenario of the \textit{GenColor} in tasks such as identifying colors for given concepts and utilizing the identified colors to color clipart, highlighting its potential to support designers in real-world scenarios (Sect.~\ref{ssec:appication-scenario}).
A gallery of color-concept associations has been developed and will be made publicly available to facilitate research in this direction (Sect.~\ref{ssec:gallery}).

In summary, our work makes the following contributions:
\begin{itemize}
    \item We identify and distill design considerations of associating color with concepts, emphasizing the need for context-dependent concepts and primary-accent color composition.

    \item We propose \textit{GenColor}, a novel framework that leverages generative models for obtaining representative image samples and enabling robust post-process for handling image variation.

    \item We collect a coloring dataset from professional designers, and the effectiveness of the \textit{GenColor} is evaluated through quantitative comparison with existing query-based approaches, showing the potential of the \textit{GenColor} in supporting designers in real-world design tasks.
\end{itemize}

\section{Related Work}
\label{sec:related}
\subsection{Color in Visual Design}

Color is a fundamental element in visual design, with applications in various domains, including graphic design~\cite{kim2014perceptually,qiu2022intelligent,shi2023stijl}, visualization design~\cite{setlur2016linguistic,shi2022colorcook}, and interior design~\cite{lin2022c3,hou2024c2ideas}.
Effectively using colors is essential for conveying information, evoking emotions, and enhancing appeal.
\re{Various previous literatures have focused on color perception, including contrast~\cite{sandnes2021inverse}, harmony~\cite{shen2000color, lin2022c3}, and psychological impacts~\cite{ou2004study, kurt2014effects}. 
Adobe has integrated these perception guidances into their Spectrum color system to enhance usability and accessibility~\cite{adobeSpectrumColors}.}
Besides, color semantics, which refers to the way people associate colors with specific concepts, plays a crucial role in enhancing communication beyond effectiveness and aesthetics~\cite{lin2013selecting,setlur2016linguistic,shi2022colorcook}, such as \q{red} with \q{apples} or \q{blue} with the \q{sky}.
Thoughtful application of color semantics can enhance the conveyance of meaning and emotion, enhancing communication beyond mere aesthetics.
For example, using semantic-resonant colors can improve cognitive performance in interpreting information visualization, allowing users to understand data categories without constantly referencing a color legend~\cite{lin2013selecting}.
Moreover, these colors can convey abstract concepts like style or theme, aiding in the communication of the overall ambiance of a design.
As such, color semantic mapping is a prevalent strategy in various design disciplines like industrial design~\cite{kobayashi1981aim, lee2006development} and dashboard design~\cite{shi2022colorcook}.

Conventional approaches to associating color semantics rely on designers' expertise and personal experience, which may be influenced by cultural background, personal preferences, and past experiences~\cite{lee2015effects}.
\re{However, clients may have different preferences for color-concept associations compared to designers~\cite{annie2018color}.
There is a growing need to provide designers with references on color-concept associations~\cite{hegemann2024palette}.}
To address this gap, computational approaches for mapping color-concept associations have emerged, including crowdsourcing efforts like the XKCD color survey~\cite{xkcd2010color} and data mining from large image or language databases~\cite{setlur2016linguistic}.
In this study, we introduce a novel approach that leverages generative models, instead of searching for references, to investigate color-concept associations.
Drawing inspiration from generative models trained on extensive real-world image datasets, our approach can capture various color concepts, align well with designer perceptions, and offer designers reference colors for diverse concepts.

\subsection{Computational Approaches for Color-Concept Association}
Associating colors with concepts traditionally relies on human ratings, where participants are tasked with selecting colors that best match given concepts~\cite{wright1962meanings,ou2004study}, providing names for given colors~\cite{xkcd2010color}, or assessing the strength of these  associations~\cite{tham2020systematic,schloss2018color,jonauskaite2019machine}.
While effective, these methods are time-consuming, labor-intensive, and limited in scope, covering only a small number of concepts and colors.
To address these limitations, researchers have developed automated methods that extract color-concept associations through statistical analysis of language and image databases.
Among them, linguistic-based approaches use the frequency of basic color terms (\eg, \q{red} and \q{blue}) in language as indicators of color-concept associations~\cite{setlur2016linguistic}.
These linguistic methods are often combined with image-based approaches to provide more comprehensive color information~\cite{lindner2012color,lin2013selecting}.
For example, Setlur and Maureen~\cite{setlur2016linguistic} developed a two-stage framework that leverages \re{N-gram}~\cite{michel2011quantitative} and Google Image Search to map a word or phrase to a color.
The method pairs a given term with each of the eleven Berlin \& Kay basic color terms~\cite{berlin1991basic} and calculates their co-occurrences based on querying them in the Google \re{N-gram} corpus\footnote{\url{https://books.google.com/ngrams/}}.
Once the highly related basic color terms are identified (\eg, associating "red" and "green" to "apple"), a triple-tuple such as ("apple", "red", "clipart") is submitted to Google Image Search API to retrieve images and extract the dominant colors from these reference images.

However, these language or image database-based approaches face significant challenges.
Language databases, such as Google N-gram, fall short in covering the combinations of context-dependent concepts and color terms, thus compromising the accurate justification of their association.
For example, \q{quiet forest} and \q{polluted sky} are common descriptions, while their combinations with basic color items are inaccessible in the \re{N-gram} corpus.
As the complexity of concept descriptions increases, the limitations of \re{N-gram-based} approaches become more pronounced~\cite{rathore2020estimating,clark2013statistical}.
On the other hand, the search-based reference images are unstable and involve varying image conditions, making further color extraction challenging~\cite{hu2023self}.
For example, the searched images may contain co-occurring concepts irrelevant to the target concept, leading to misleading color-concept association.
To mitigate the issues, existing work limits the search corpus to simple clipart~\cite{setlur2016linguistic, lin2013selecting} or only derives color histogram rather than dominant colors~\cite{hu2023self}.
However, these constraints inherently limit the application scope.
In this work, we introduce a generative approach to color-concept association.
We leverage the powerful generative capabilities and controllability of diffusion models to flexibly synthesize contexts and concepts, and control image conditions (\eg, fidelity and illumination), thereby overcoming the limitations of traditional language and image database methods.

\subsection{Generative Models as Tools for Data Mining}
Generative diffusion models have demonstrated powerful capabilities in producing high-fidelity images from prompts~\cite{rombach2022high, saharia2022photorealistic}.
Research has further enhanced their applicability in scenarios like controllable generation~\cite{zhang2023adding}, image editing~\cite{kawar2023imagic, brooks2023instructpix2pix} and inpainting~\cite{lugmayr2022repaint}.
Building on these advancements, the HCI community has explored use cases that utilize generative models for human-AI co-creation~\cite{verheijden2023collaborative, fan2024contextcam}, such as typographical design~\cite{xiao2024typedance} and product personalization~\cite{shi2024personalizing}.
In contrast to previous works focusing on creative design, our work employs generative models as \textbf{a data mining tool to uncover real-world visual patterns, beyond mere image synthesis}.
Our motivation stems from the ability of generative models to capture complex patterns and variations in the training data, thus generating new, realistic samples that reflect the characteristics of the original data.
Recent studies have employed pretrained diffusion models to generate synthetic data for training models, thereby improving their performance on classical computer vision tasks such as image classification~\cite{sariyildiz2023fake, azizi2023synthetic} and image segmentation~\cite{nguyen2024dataset}.
For example, Sariyildiz \etal~\cite{sariyildiz2023fake} leverage Stable Diffusion~\cite{rombach2022high} to generate synthetic ImageNet~\cite{deng2009imagenet} clones, \ie, datasets with synthetic images for the ImageNet classes, using class names as prompts.
Close to our work, researchers have trained generative models (\eg, generative adversarial networks~\cite{goodfellow2014generative}) on faces~\cite{chen2023s} and cars~\cite{dalens2019bilinear} image datasets to analyze the evolution of their characteristics over history.

Leveraging diffusion models to generate data for training and analysis in vision-centric tasks represents a novel paradigm.
The approach can reduce manual labor and mitigate potential data collection issues, such as poor image quality and mislabeled data~\cite{muller2019identifying}.
To efficiently explore visual patterns of specific interest from the large visual representation space of pre-trained models, fine-tuning~\cite{ruiz2023dreambooth} and prompt engineering~\cite{hao2024optimizing} are often required to achieve precise control over the generated content.
For example, Siglidis \etal~\cite{ioannis2024diffusion} introduced an approach that fine-tunes diffusion models on specific datasets to facilitate the generation of image samples that align closely with their labels.
However, fine-tuning is more appropriate for closed-set concepts (\eg, classification tasks with finite concept labels), while our target is open-set color-concept associations.
To achieve a precise analysis of the intended visual concepts, we introduce a robust post-processing pipeline along with prompt tuning that achieves flexible control.
Specifically, we integrate open-set detection~\cite{liu2023grounding} and promptable segmentation~\cite{kirillov2023segment} to precisely segment concept-relevant regions within the image and then conduct color association extraction for primary and accent colors.
\section{Formative Study}
\label{sec:preliminary}

\subsection{Study Design}
To gain insights into how professional designers apply color-concept associations, we conducted semi-structured interviews with four experienced designers, each lasting around 45 minutes.
The participants brought diverse experiences: E1 had over 10 years of experience in visual design, E2 had 6-10 years, and E3 and E4 had 3-5 years of experience.
Their expertise spanned graphic design, UI design, game art design, and product design, offering valuable insights into the role of color in different design contexts.
Two participants had formal academic backgrounds in design-related fields, while the others combined academic knowledge with industry practice, offering a well-rounded perspective on how theoretical concepts are applied in professional settings.

During the interviews, we explored the following topics:
(1) the workflows designers use when selecting colors for specific concepts,
(2) the guiding principles and considerations for semantic color associations and
(3) designers' expectations for \re{a supportive tool that} integrates color-concept associations into their processes.
All interviews were recorded and transcribed to facilitate a detailed analysis process.

\subsection{Designer Feedback}
In line with previous research, associating colors with specific concepts to convey meaning is a common practice in visual design~\cite{osgood1957measurement, palmer2010ecological, humphrey2019colour}.
Here, we use the example of color-concept association in clipart design, where the task is to design a clipart of \q{mountain}, as illustrated in Figure~\ref{fig:usage_scenario}.
Given different contexts (\q{pristine} \vs \q{polluted}), the anticipated designs and corresponding colors are different.
In the following, we use this example and summarize the designers' feedback into key considerations in design practice and workflow.

\subsubsection{Considerations in Design Practice}
In color-concept association for visual design, the concept is \emph{context-dependent}, and the choice of color typically involves a composition of both \emph{primary and accent colors}.

\begin{figure*}[t]
      \centering
      \includegraphics[width=0.985\linewidth]{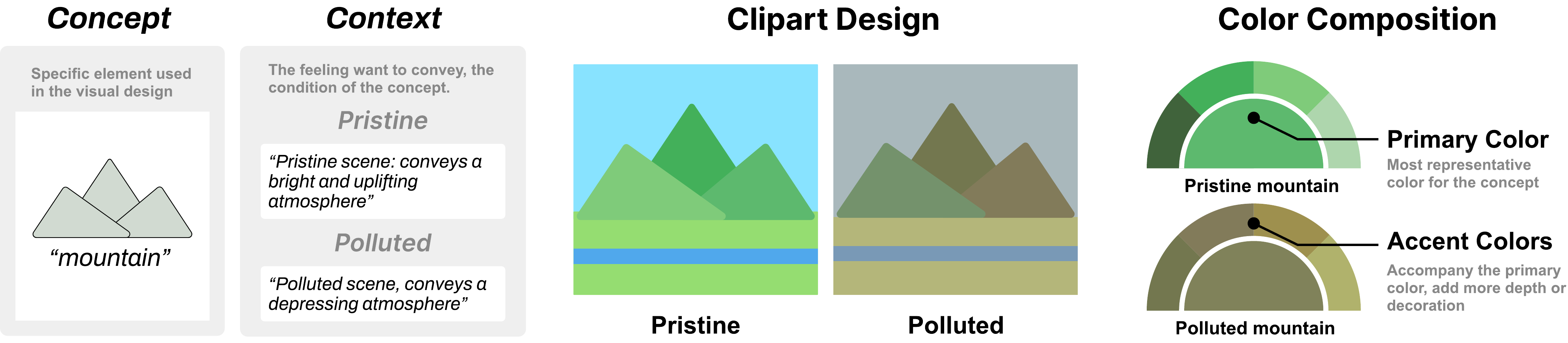}
      \caption{Usage scenario for designing a clipart to raise awareness about mountain pollution. (a) Illustration of the design elements, including the concept (\q{mountain}) and the context (\q{pristine} \vs \q{pollution}). (b) Examples of clipart design: the clipart for \q{pristine mountain} uses bright green, whilst the one for \q{polluted mountain} uses grayish green. (c) Primary-accent color composition is used by designers, with the primary color representing the concept and the context, while accent colors provide depth and decoration.}
      \Description{Usage scenario for designing a clipart to raise awareness about mountain pollution. (a) Illustration of the design elements, including the concept (\q{mountain}) and the context (\q{pristine} \vs \q{pollution}). (b) Examples of clipart design: the clipart for \q{pristine mountain} uses bright green, whilst the one for \q{polluted mountain} uses grayish green. (c) Primary-accent color composition is used by designers, with the primary color representing the concept and the context, while accent colors provide depth and decoration.}
      \label{fig:usage_scenario}
\end{figure*}
\begin{itemize}
      \item \emph{Context-dependent concept.}
            Designers often associate colors with various meanings~\cite{chen2021investigation}, such as emotions, styles, cultures, or specific items.
            We refine this broad notion of \q{meaning} into two distinct components: \textit{concept} and \textit{context}.
            The \textit{concept} refers to a concrete design element, such as a mountain, tree, or car, while the \textit{context} reflects the abstract emotional tone or condition to be conveyed~\cite{jahanian2017colors, hegemann2024palette}.
            For example, in a clipart designed to raise awareness about environmental pollution, the \q{mountain} serves as the concept, while its \q{clean} or \q{polluted} state represents the context, conveying a sense of environmental purity or degradation, as shown in Figure \ref{fig:usage_scenario}.
            Bright greens can evoke an uplifting atmosphere of a pristine mountain, while grayish greens can depict the degradation of a polluted mountain.
            This context-dependent color choice adjusts the design's mood without altering the core concept.
            Proper colors align messages with both concept and context.
            A mismatch in color-concept association, such as using bright green for a polluted mountain, could confuse viewers and weaken the intended message.
      \item \emph{Primary-accent color composition.}
            Color composition rules are considered a fundamental principle in visual design.
            When associating with concepts, the primary color is closely associated with the concept and the context being depicted.
            It typically occupies the largest area in the design, ensuring that the concept remains central and easily recognizable.
            Accent colors complement the primary color to create a balanced and meaningful visual representation.
            They provide decorative or contextual elements to support the overall design.
            In the example, the primary color is bright green for \q{pristine mountain} and grayish green for \q{polluted mountain}.
            The accent colors, such as darker and lighter shades of green, are used to suggest distance, depth, and shading.

\end{itemize}

\noindent
\textbf{Concept and Context-Dependent Color Selection.}
All participants emphasized that the concept is the primary consideration when selecting colors to convey specific meanings.
There are some common color-concept associations used in design, such as blue for the sky, green for mountains, and brown for soil.
Besides, they also pointed out that color selection is influenced by factors beyond the concept itself.
These factors can be considered as contexts, such as feelings, styles, and target audiences.
While all participants mentioned feelings, they acknowledged that feelings can be vague.
For instance, E1 described designing a poster about lime, explaining that instead of using the saturated green of the lime peel, they would opt for a fresher green to convey a \q{refreshing taste.}
E1 elaborated, \q{If I look at the context, I would definitely consider its sour feeling, and I think it must be related to the green and yellow color.}
E3 also highlighted that sometimes the customer's requirements are vague, such as \q{changing the color to be more active.}
Regarding style, E2 pointed out that color usage differs between flat and realistic designs, with flat designs typically using more saturated colors.
Additionally, participants may consider the audience of their designs.
For instance, E3 noted that designs aimed at children tend to use brighter, more vivid colors, while business designs employ more subdued tones to convey professionalism.
There is a need for a flexible approach that can adapt to these contextual factors, providing designers with a more nuanced and tailored color recommendation.

\noindent
\textbf{Intuitive Representation for Color Composition.}
Designers consistently expressed the need for a well-defined color composition, with a highlight on the primary color and some accent colors.
E2 gave an example of food packaging for different flavors, explaining, \q{The color should be representative of the flavor, such as tomato-flavored potato chips are always red. So when buyers see it, they can quickly know what flavor it is.}
E4 also preferred starting with a primary color tied to the concept, complemented by other colors working together in the composition.
Participants were critical of existing color representations of possibility distributions, with E1 describing them as \q{not intuitive} and \q{too many colors unrelated to the concept.}
A palette containing around five colors, as E1 mentioned, was deemed sufficient, similar to those provided by design tools.
While the current color palette is insufficient, as usually treated equally with all colors.
They mentioned it would be helpful to make the primary color salient, in the center, with accent colors surrounding it, to provide a clear visual hierarchy.
Besides, they also suggested that the color ratio should be provided, even if not strictly followed, to give designers a sense of the relative importance of each color, which aligns with previous research~\cite{shi2023stijl}.

\subsubsection{Design Workflow}
\label{sssec:design-practice}

The practice of color-concept association can generally be categorized into conventional and computational approaches.

\begin{itemize}
      \item \textit{Conventional approach}.
            Conventionally, designers begin by gathering references to identify color associations with specific concepts.
            They search for relevant images, photography, or game scenes that effectively convey the intended concept, using image searching engines like Google Images\footnote{\url{https://images.google.com/}} or Pinterest\footnote{\url{https://www.pinterest.com/}}.
            Many designers habitually collect references, often maintaining large and high-quality personal libraries for inspiration.
            From these image references, designers draw inspiration for primary colors and iteratively identify accent colors.
            Various tools, such as eyedroppers for color extraction and color wheels for modifying color compositions, are utilized to aid this process.

      \item \textit{Computational approach}.
            Recent advancements in linguistic analysis and image processing have boosted the development of computational approaches for color-concept association.
            This approach typically involves several steps: identifying relevant concepts in text corpora (\eg, Google Ngrams\footnote{\url{https://books.google.com/ngrams/}}), finding reference images from an image dataset, and extracting colors from those reference images.
            These processes can all be automated to a large extent, leveraging machine learning techniques.
            The CIELAB color space~\cite{commission1978recommendations} is often used in the color extraction process due to its computational simplicity and relatively perceptually uniform properties.

\end{itemize}

\begin{figure*}[t]
      \centering
      \includegraphics[width=0.985\textwidth]{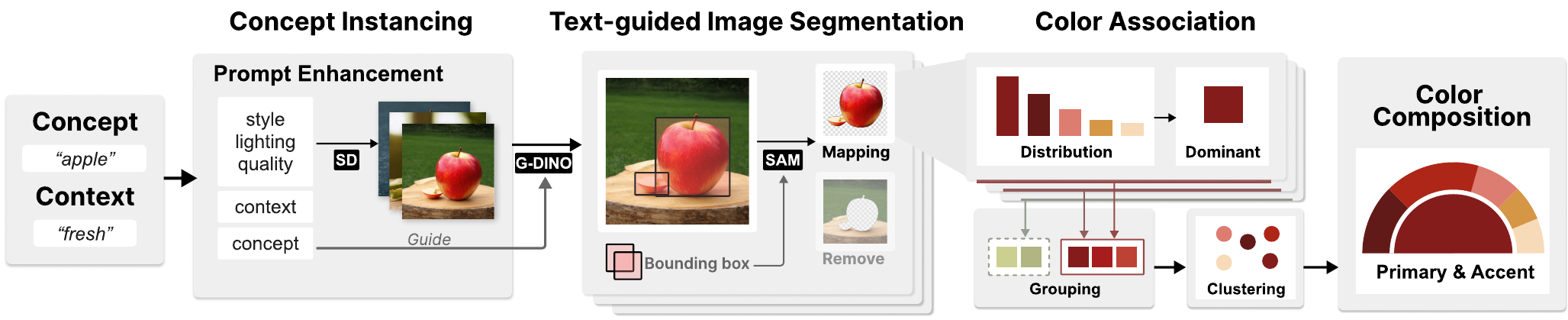}
      \caption{Overview of the \textit{GenColor} framework. The framework includes three stages: Concept Instancing for generating representative image samples, Text-guided Image Segmentation for identifying relevant regions, and Color Association for extracting the primary-accent color composition.}
      \Description{Overview of the \textit{GenColor} framework. The framework includes three stages: Concept Instancing for generating representative image samples, Text-guided Image Segmentation for identifying relevant regions, and Color Association for extracting the primary-accent color composition.}
      \label{fig:pipe}
  \end{figure*}

\noindent
\textbf{Universal and Objective Color References.}
In the conventional approach, we found that the process of associating colors with concepts is varies among designers.
Each step relies heavily on intuition and is influenced by personal preferences.
While general principles such as color composition rules (E3) and color psychology (E1 and E4) provide guidelines, the final color choice remains subjective.
Personal experience and painting styles also play a role.
For instance, E4 prefers Morandi colors with lower saturation, while E1 favors brighter colors.
This variability extends to how designers approach different concepts, depending on their familiarity and personal feelings.
E3 mentioned that what design school actually teaches is aesthetics, which is \q{when you see something good, recognize the feeling, and know how to recreate it}.
While courses may teach step-by-step techniques, \re{effective design work also benefits from objective references that help designers understand the public's perception.}
As E3 explained, \q{Sometimes they (designers) focus only on what they want to express, but what others perceive may be different from their intent, and the result is not what their customer wants.}
Participants believe there is a need for an objective reference to provide a useful benchmark, even if it isn't directly used in the final design.
Such a reference, aligned with public perception, would help guide designers toward more universally understood choices, mentioned by E3 as \q{move closer to objective data.}
This could also benefit novice designers, as E1 noted, who may not yet be familiar with or sensitive to certain design concepts and need a starting point for color selection.

\noindent
\textbf{Flexible and Robust Image-Based Computational Methods.}
The computational approach can be limited by the number and quality of images available in the dataset.
While designers perceive computational methods as more objective than conventional approaches, the results are often unsatisfactory due to the limitations of image referencing.
E2 mentioned that the retrieved images are not always relevant or fail to capture the desired atmosphere.
This limitation is attributed to the search engine's algorithm's inability to consider context.
For example, while there may be numerous image references for a generic concept like \q{mountain}, there are often fewer relevant references for a more specific concept like \q{polluted mountain}.
Moreover, the search results may not use color to reflect the context of \q{polluted}; they might simply add visual elements like garbage without modifying the color to convey degradation.
Such retrieved image results are not helpful for finding color references that align with the intended concept and context.
Additionally, the retrieved images are often of low quality, with cluttered backgrounds or inconsistent lighting, further complicating the task of identifying suitable color references.

\subsection{Design Goals}

Facing the gap between designer requirements and existing approaches, we aim to address these issues by leveraging generative models for color-concept association.
Our work is inspired by contemporary \re{T2I} models, which can resemble visual patterns from large-scale real-world data.
Based on insights from the formative study, we identified the following design goals:

\begin{enumerate}
      \item \textbf{Provide Justified and Objective Color References.}
            The framework should be both justifiable and transparent, offering a clear rationale for each suggested color, ensuring designers feel confident about the result.
            To align with designers' practice of extracting colors from image references, the framework shall compute colors from image references, rather than directly give a coloring result.
            Additionally, these color references should be derived from large-scale datasets, helping to minimize personal biases and providing objective guidance in the color selection process.

      \item \textbf{Maximize Feasibility and Flexibility in Concepts and Context.}
            The framework should accommodate a wide range of concepts and contexts encountered in design.
            It must dynamically adapt to various conceptual entities and contextual factors, such as emotional tone, artistic style, and target audience.
            The generative approach should flexibly adjust to these variations, ensuring the generated colors align with the core concept while capturing subtle contextual nuances, offering a tailored response to each unique design scenario.

            \item\textbf{Facilitate Robustness and Effectiveness in Associating Primary-accent Color.}
            The framework should establish robust color-concept associations by providing a primary-accent color composition that reflects the concept and context.
            It should identify semantically related regions in reference images and minimize noise during the process.
            The color composition should consist of a primary color that closely reflects the concept, with accent colors providing additional depth and variety while still maintaining relevance to the overall design.
\end{enumerate}

\section{Generative Approach for Color-Concept Association}
\label{sec:gencolor}
This section presents \emph{GenColor}, an innovative framework that automatically associates a given concept with primary-accent color composition mined from images by generative models.
As depicted in Figure~\ref{fig:pipe}, the framework comprises three stages:
(1) \textit{Conceptual Instancing} to generate representative images samples (Sect.~\ref{ssec:conceptual-instancing}),
(2) \textit{Text-guided Image Segmentation} to identify conceptural-relevant image regions for color analysis (Sect.~\ref{ssec:image-segmentation}), and
(3) \textit{Color Association} to extract primary-accent color compositions (Sect.~\ref{ssec:color-association}).
Such a framework mirrors the professional design process of sourcing references and selecting visually appealing concept colors, as discussed in \textit{Design Practice} (Sect.~\ref{sssec:design-practice}).

\begin{figure*}[t]
    \centering
    \includegraphics[width=0.985\textwidth]{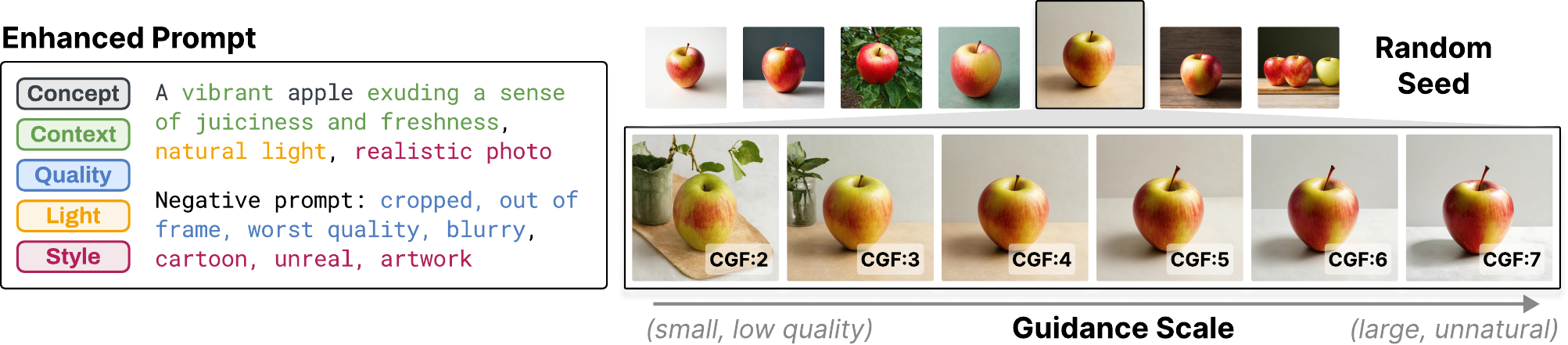}
    \caption{\re{Prompt design and parameter setting in the Conceptual Instancing stage. The prompt is refined based on the core principles of concept, context, style, lighting, and quality control. The guidance scale and seed are adjusted to control image quality and diversity.
    }}
    \Description{Prompt design and parameter setting in the Conceptual Instancing stage. The prompt is refined based on the core principles of concept, context, style, lighting, and quality control. The guidance scale and seed are adjusted to control image quality and diversity.}
    \label{fig:concept-instancing}
\end{figure*}

\subsection{Conceptual Instancing}
\label{ssec:conceptual-instancing}
The initial stage of our framework is dedicated to leveraging diffusion models, specifically Stable Diffusion 3, to generate images that accurately represent a given concept.
Given the critical role of prompts and hyperparameters in the image generation process and the potential risk of producing duplicate images, this stage involves two primary steps: a universal prompt refining strategy and a generation strategy varying the hyperparameters, ensuring the diffusion model to generate images that reflect real-world color-concept associations.

\subsubsection{Enhancing the Concept Prompt}
Prompt design is critical for the quality and relevance of generated images, as highlighted in previous works~\cite{liu2022design,dalleprompt2022guy}.
An effective prompt should strike a balance between specificity and flexibility, ensuring accurate visual interpretation while avoiding unnecessary artifacts or distortions.
To ensure realistic and consistent color reproduction, we must avoid introducing unnecessary color variance that diverges from the intended concepts and contexts.
At the same time, developing a flexible prompt template enables more effective control over image conditions, allowing for the seamless synthesis of concepts across various contexts, as shown in Figure \ref{fig:concept-instancing} (left).

Specifically, we focus on the following core principles when refining our prompts:
\begin{itemize}
    \item \textbf{Concept}: Clearly define the core concept to be visually represented. Not introducing extra elements or irrelevant objects ensures that the generated images are focused and accurate. 
    \item \textbf{Context}: Context refers to the conditional or emotional state of a concept. For example, \q{ripe banana} versus \q{unripe banana}, or \q{sunny sky} versus \q{stormy sky}. 
    By providing context, we guide the model to generate images that accurately capture the intended mood or setting. 
    It is important to avoid descriptions that directly reference specific color tones, as this could introduce bias into the model.
    \item \textbf{Style}:
          Style should be defined based on the intended use of the image.
          We differentiate between two common categories: \q{realistic photos} and \q{clipart.}
          However, terms like \q{clipart} can be misleading to the model, as Stable Diffusion may misinterpret it. To ensure clarity, we recommend using phrases like "colored flat design" for simpler illustrative styles and "realistic photo" for high-detail, lifelike outputs.
          By setting clear stylistic expectations, we can mitigate unwanted style mismatches, such as cartoonish or 3D elements when realism or flatness is intended.
    \item \textbf{Lighting}: Lighting should be carefully controlled to maintain realism and color consistency across images.
          Natural lighting is generally recommended avoiding distortions from artificial light sources, which can introduce color shifts or unnatural shading~\cite{witteveen2022investigating}.
    \item \textbf{Quality Control}: Ensuring the generation of high-quality images involves a combination of carefully chosen parameters and the use of negative prompts.
          Negative prompts allow us to exclude unwanted characteristics—such as overly cropped images or incorrect styles~\cite{ban2024understanding}. For example, when generating realistic photos, we would specify in the negative prompt to avoid cartoon or 3D styles, while for flat design, we would exclude realistic 3D rendering styles. This targeted exclusion helps ensure that the generated objects are fully within the frame and match the intended quality.
\end{itemize}

\begin{figure*}[t]
    \centering
    \includegraphics[width=0.985\textwidth]{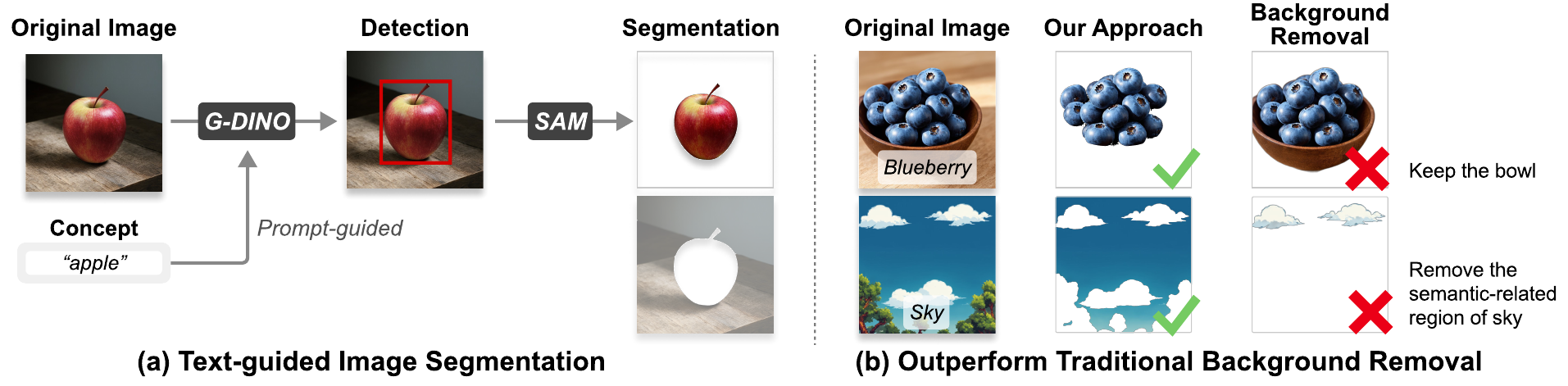}
    \caption{The text-guided image segmentation process. (a) The pipeline utilizes G-DINO for prompt-guided detection and SAM for concept-based segmentation. (b) Comparison of traditional background removal and text-guided segmentation.}
    \Description{The text-guided image segmentation process. (a) The pipeline utilizes G-DINO for prompt-guided detection and SAM for concept-based segmentation. (b) Comparison of traditional background removal and text-guided segmentation.}
    \label{fig:segmentation}
\end{figure*}

\subsubsection{Image Generation with Stable Diffusion}
After refining the concept prompts, we proceed to the image generation phase using the Stable Diffusion 3 model~\cite{rombach2022high}. 
This model is selected for its ability to produce detailed and visually appealing images that align well with specified prompts, effectively reflecting real-world color associations.
To introduce variability and creativity in the generated samples, we adjust certain parameters within the model:
\begin{itemize}
    \item \textbf{Guidance Scale}: This parameter manages the balance between strict adherence to the prompt and creative exploration.
    \re{
    Randomizing the guidance scale within an appropriate range helps generate diverse images. 
    Extremely low values can lead to distortions or incomplete images, while high values can result in unnatural contrasts (see Figure~\ref{fig:concept-instancing} (right)).
    Based on Stable Diffusion recommendations and our empirical adjustments, we set the value to vary randomly between 3 and 6.}

    \item \textbf{Seed}: Randomizing the seed parameter ensures that each image generation is unique, even when the prompt remains consistent. This randomness is crucial for producing diverse images and avoiding repetitive outputs.
\end{itemize}

We generate a set of 50 images for each concept, each with a resolution of $1024\times1024$ pixels.
\re{
The quantity of images is chosen following previous literature on obtaining color-concept association \cite{rathore2020estimating,hu2023self} to provide a diverse and comprehensive representation of the concept.
}
This resolution is chosen to provide a high level of detail, which is essential for comprehensive color analysis in subsequent stages.
Additionally, generating images at a higher resolution increases the likelihood that the model will place the object in the center of the image, resulting in more consistent and varied compositions.
By carefully adjusting these parameters, we ensure that the generated images adhere to the specified concepts and exhibit a wide range of variations, enhancing the overall quality and diversity of the image set.

\subsection{Text-guided Image Segmentation}
\label{ssec:image-segmentation}
After obtaining representative images of a specified concept, accurately identifying the concept-related regions is crucial for further precise color information extraction.
Previous query-based approaches consider unstable search results and varying image conditions, thus restricting the image type to clipart~\cite{setlur2016linguistic, lin2013selecting} or using background removal techniques to obtain image segments~\cite{hu2023self}.
However, traditional background removal has several drawbacks.
One major limitation is the insensitivity to specified concepts.
Background removal techniques may retain irrelevant co-occurring elements, leading to misleading color-concept associations.
For example, in the \q{blueberry} case, background removal will still retain unrelated elements (the bowl contains the blueberry), making subsequent color extraction difficult, as shown in Figure~\ref{fig:segmentation} (b).
Another issue is that common concepts, such as "sea" and "sky," often appear as part of the image background.
Background removal techniques might eliminate these elements, which are common in visual design.
For example, Figure~\ref{fig:segmentation} (b) presents a case of "sky," where the background removal technique retains the foreground clouds while removing the "sky" we are interested in.
To address the issues, we employ a text-guided image segmentation approach that seamlessly integrates with previous text-guided image generation, as depicted in Figure~\ref{fig:segmentation}.
This integration enables the effective and efficient segmentation of concept-relevant regions in the generated images, thereby enhancing the accuracy of our analysis.

We aim to use specified concepts as text prompts to get the corresponding image segment directly.
However, in computer vision, determining the mask in an image that matches the region described by text remains challenging.
To effectively address this open-set segmentation challenge~\cite{ren2024grounded}, we employ an ensemble approach, dividing it into two main components: open-set detection~\cite{liu2023grounding} and promptable segmentation~\cite{kirillov2023segment}.
Figure~\ref{fig:segmentation} (a) illustrates the process, which begins with GroundingDINO~\cite{liu2023grounding}, an open-set object detector that uses textual input to generate bounding boxes around regions of interest within the image.
To refine these bounding boxes, we apply Non-Maximum Suppression (NMS) to eliminate overlapping regions, retaining only the most confident and relevant detections.
Subsequently, the annotated boxes obtained through GroundingDINO serve as the box prompts for the Segment Anything Model (SAM)~\cite{kirillov2023segment}, which generates precise mask annotations for the identified regions.
After segmentation, all segmented regions are merged into a single mask, representing the concept-relevant areas within the image.
By leveraging the capabilities of these two robust expert models, we can more effectively accomplish open-set detection and segmentation tasks, thereby improving the accuracy of color-concept association analysis.

\subsection{Color Association}
\label{ssec:color-association}
\begin{figure*}[t]
    \centering
    \includegraphics[width=0.9\linewidth]{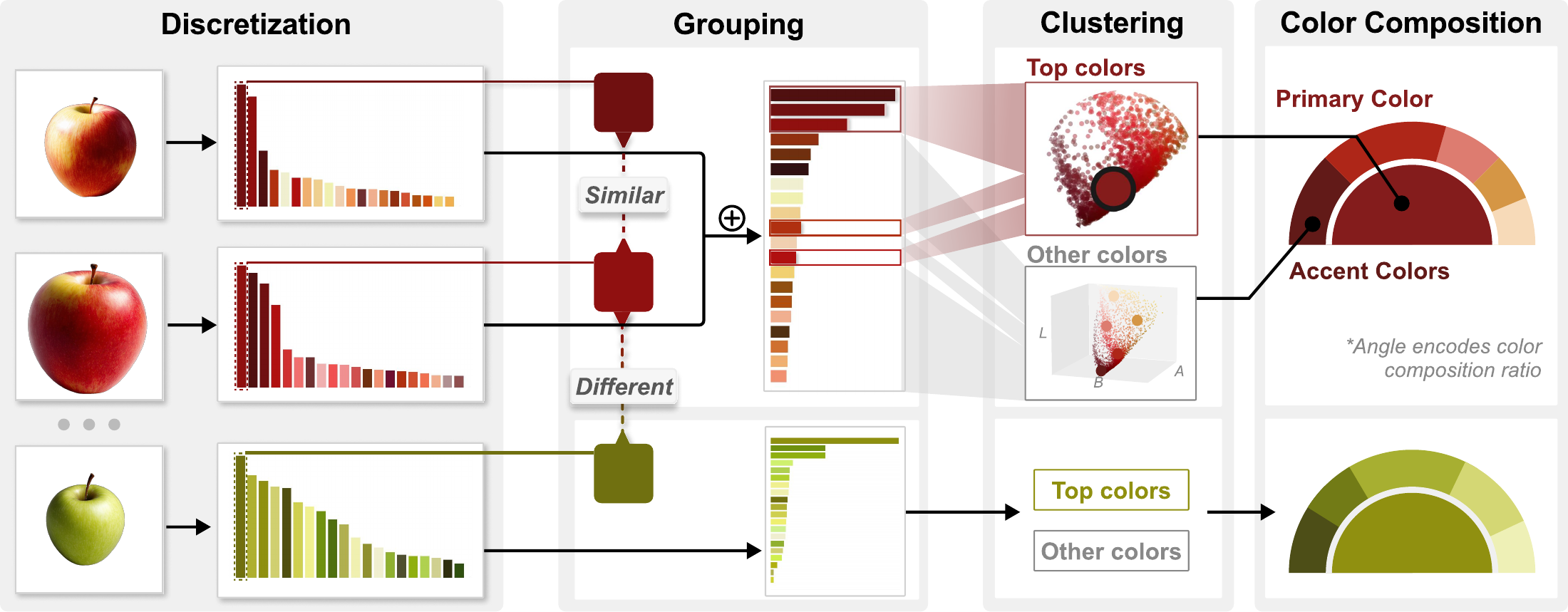}
    \caption{The process of extracting color information from segmented regions of generated images to derive a primary-accent color composition for a given concept. The framework includes color discretization, grouping similar colors into same groups, and clustering to identify the primary and accent colors for the final color composition.}
    \Description{The process of extracting color information from segmented regions of generated images to derive a primary-accent color composition for a given concept. The framework includes color discretization, grouping similar colors into same groups, and clustering to identify the primary and accent colors for the final color composition.}
    \label{fig:color-extraction}
\end{figure*}

The final stage of our framework focuses on extracting color information from the segmented regions of generated images and deriving a primary-accent representative color composition for given concepts.
The objective is to identify the primary and accent colors that best represent the concept, ensuring that the color composition is both accurate and meaningful for design applications.
Figure~\ref{fig:color-extraction} illustrates the color association pipeline.
Our pipeline begins with color discretization, where the color values within the segmented regions are mapped into a discretized $16\times16\times16$ RGB color space.
This step reduces the complexity of the color data by categorizing close color values into defined bins, simplifying identifying the most frequent colors within each image.
The color corresponding to the bin with the highest count is then selected as the dominant color at the image level, such as \raisebox{-.2\height}{\includegraphics[width=0.32cm]{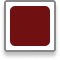}} \raisebox{-.2\height}{\includegraphics[width=0.32cm]{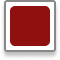}}
\raisebox{-.2\height}{\includegraphics[width=0.32cm]{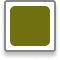}} in Figure~\ref{fig:color-extraction}.

Considering that a concept can be associated with multiple reasonable colors across different images, it is essential to group colors appropriately.
\re{
Our approach aims to balance conciseness and diversity in color representation, thereby providing an effective color abstraction.
We merge similar colors, while still allowing diversity at the color palette level.}
Specifically, displaying multiple reasonable colors together should be avoided, as they may not typically appear simultaneously in a design.
For instance, an \q{apple} is represented by different dominant colors like red or green, but it would be unusual to see both colors in one apple.
\re{
To preserve diversity within the same color category, we distinguish between different shades of similar colors. 
Take \q{mountains}, for instance; they can appear in deep green or deep blue-green tones. 
Although these shades are somewhat similar, separating them into distinct groups ensures a varied and comprehensive color palette and provides diverse references for designers.
}
Thus, we conduct image grouping based on their dominant colors.
To address this, we group images by calculating CIEDE2000 color distances in the CIELAB space and applying a threshold of 12.
Such a threshold allows for noticeable differences between colors, such as variations within the same color category (\eg, different shades of red), while still considering them part of the same color group.
This ensures that similar colors are clustered together while distinctly different colors remain separate.
\re{After grouping, we sort the color groups by the number of images they contain and select the top five most frequent color groups. 
To ensure the robustness of the pipeline, we exclude any color groups that contain fewer than three images.}
After grouping, we aggregate the color bin values of each image in the same group.

Similar to image-level color extraction, we apply color discretization to the aggregated color bins of each image group.
The top colors are considered the dominant color for the group.
To deliver the color composition of each image group, we separate their aggregated color bins into two classes: \textit{top colors} and \textit{other colors}.
Specifically, using the CIEDE2000 color distance threshold $\leq7$, commonly recognized as the range distinguishable by the human eye~\cite{szafir2017modeling}, we filter the colors surrounding the group-level dominant color as the \textit{top colors}.
Colors not falling within this threshold are categorized as \textit{other colors}.
Then, we set the centroid of \textit{top colors} as the primary color.
For accent colors, we apply $k$-means clustering to compute the typical colors for the concept across all images, where we set $k$ to 5 following previous color palette research~\cite{chang2015palette}.
This grouping-then-clustering approach ensures the full exploration of color-concept association and the delivery of the primary and accent colors to represent the concept, thus providing a coherent and comprehensive color composition for design purposes.

For the final color composition, we design an intuitive visual representation that combines the primary and accent colors.
Inspired by the traditional painting palette, we create a radial glyph, placing the primary color in the center and the accent colors around it, as shown in Figure~\ref{fig:color-extraction}.
The central position of the primary color emphasizes its importance, allowing designers to quickly identify the main color associated with the concept.
Accent colors with greater variance are positioned on the outer ring, helping designers explore a broader range of complementary colors.
The angular range of each accent color represents its proportion within the color palette, reflecting its frequency as determined by the clustering algorithm. 
This visual representation provides a clear and intuitive way to present the color composition of a concept, aiding designers in selecting appropriate colors for their design tasks.

\section{Evaluation and Application}
\label{sec:evaluation}
This section presents a comprehensive evaluation of \textit{GenColor}, demonstrating that it generates more representative and user-preferred colors compared to previous computational methods.
To quantitatively assess the different methods in color space using color distances, we first gather a baseline dataset from designers (Sect.~\ref{ssec:designer-baseline}).
Then, we compare different approaches using color difference comparison (Sect.~\ref{sssec:color-differenece-comparison}) and user ratings (Sect.~ \ref{sssec:user-rating}).
Furthermore, we present various downstream application scenarios of \textit{GenColor} (Sect.~\ref{ssec:appication-scenario}).
Finally, we construct a large-scale color gallery where designers can explore and retrieve relevant color compositions by performing
text-based search (Sect.~\ref{ssec:gallery}).

\begin{table*}[htb]
    \centering
    \renewcommand{\arraystretch}{1.3}
    \caption{Examples of entities classified by type and subtype.}
    \Description{Examples of entities classified by type and subtype.}
    \begin{tabular}{|l|l|l|}
        \hline
        \textbf{Type}                     & \textbf{Subtype} & \multicolumn{1}{c|}{\textbf{Entity}}                              \\ \hline
        \multirow{4}{*}{\textbf{Basic Concept}}
                                          & Fruit            & apple, banana, blueberry, cherry, grape, peach, tangerine         \\ \cline{2-3}
                                          & Vegetable        & carrot, celery, corn, eggplant, mushroom, olive, tomato           \\ \cline{2-3}
                                          & Environment      & mountains, sky, ocean, desert, grassland, forest, lake, glacier   \\ \cline{2-3}
                                          & Animal           & fox, tiger, lion, elephant, crocodile, flamingo, penguin, peacock \\ \hline
        \multirow{2}{*}{\textbf{Context}} & Conditional      & polluted, clear, pure,  at springtime, at sunset
        \\ \cline{2-3}
                                          & Emotional        & lively, quiet, peaceful, depressed, happy, sad, energetic         \\ \hline
    \end{tabular}
    \label{tab:entity_classification}
\end{table*}

\subsection{Designers' Coloring Dataset}
\label{ssec:designer-baseline}
The lack of baseline data hinders the quantitative assessment of different approaches in the color space.
Here we aim to collect a dataset that shows how professional designers associate colors with various concepts and contexts, and then use this dataset as a baseline for evaluating the effectiveness of our generative approach.

\subsubsection{Scope of the Concepts and Contexts}
We selected a diverse range of concepts and contexts commonly used in design practice, drawing from previous studies and designer interests.
The selected concepts fall into two main categories: \emph{Concept} and \emph{Context}, as summarized in Table~\ref{tab:entity_classification}.

\begin{itemize}
    \item \textbf{\emph{Concept}}:
          We primarily focused on natural elements, as they tend to exhibit more stable color associations compared to artificial concepts, which can vary more widely (\eg, color of plastic).
          Based on previous studies~\cite{setlur2016linguistic, mukherjee2022context, rathore2020estimating}, we included simple, standalone entities such as common fruits (\eg, apple, banana) and vegetables (\eg, carrot, corn), with seven items in each category.
          To broaden the scope, we also incorporated more complex elements, including animals and environmental components, adding eight items per category.
          Examples include animals like fox, tiger, and lion, and environmental elements like mountains, sky, and ocean.

    \item \textbf{\emph{Context}}:
          This category pairs concepts with different contexts that can influence their color associations.
          We considered both the conditional and emotional aspects of each concept, selecting six items in this category.
          Examples include the sky in different states (\eg, polluted \textit{vs.} clear) or scenarios evoking different feelings (\eg, quiet \textit{vs.} lively).

\end{itemize}

\subsubsection{Participants and Procedure}
We recruited 24 professional designers from online design communities, representing a diverse demographic in visual design. Participants ranged from 20 to 30 years old ($M=25.5$, $SD=2.57$), with 14 males and 10 females.
Professionally, 4 worked in industry, 11 held related degrees and were employed in academia, and 8 had experience in both.
Their backgrounds spanned graphic design, UI design, interior design, illustration, product design, and animation.
In terms of expertise, 5 had 1-2 years of experience, 10 had 3-5 years, 7 had 6-10 years, and 1 had over 10 years.
\re{This study received approval from the university's ethics committee.}
The study began with obtaining informed consent and collecting background information through a questionnaire.
\rec{All participants were explicitly informed that their work contributed to the development of a Generative AI tool, specifically in the evaluation stage, and that their data would be used solely for research purposes.}
They then received detailed instructions and completed a tutorial session featuring three example concepts. 
After confirming their understanding, they proceeded with the main task.
Participants were asked to complete a coloring task for 36 different concepts, using the most representative colors they associated with each concept.
Designers were asked to color the drawings without any reference, using around 1-5 representative colors of their choice.
They were encouraged to reflect the importance of each color by adjusting the area it occupied in their design.
To simplify the task, we provided participants with a line drawing of each concept.
The line drawings were found online and verified by an expert designer to ensure that the concepts were easily recognizable and unambiguous.
Designers were allowed to use any tools they were familiar with, and no time limits were imposed on the task.
Upon completion, designers confirmed that their chosen colors matched their understanding of each concept.
The study was conducted online, taking approximately 1-2 hours, with participants compensated around 15 USD.

\subsubsection{Designers' Coloring Dataset}
\begin{figure}[t]
    \centering
    \includegraphics[width=0.985\linewidth]{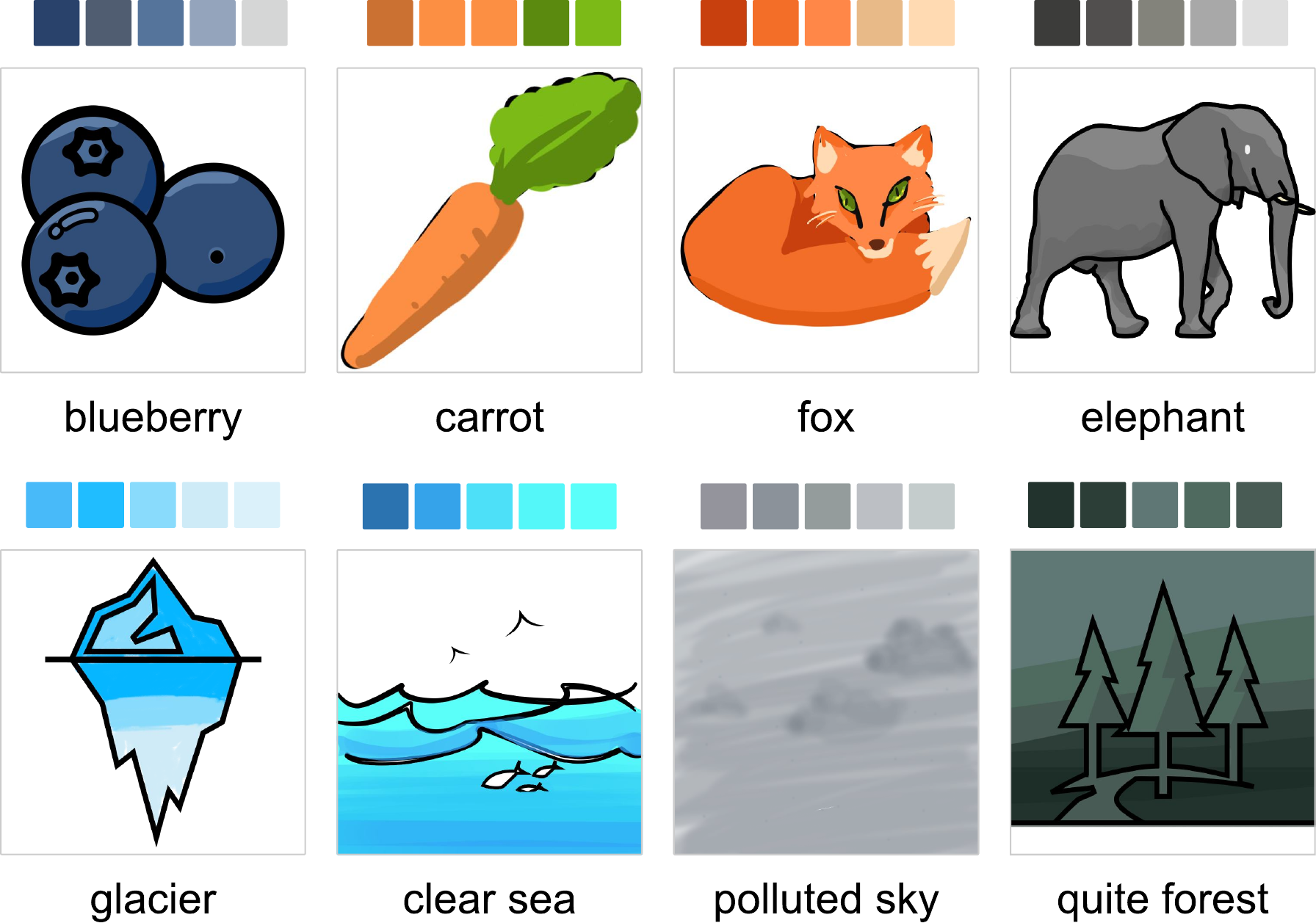}
    \caption{Examples of the dataset collected from professional designers.}
    \Description{Examples of the dataset collected from professional designers.}
    \label{fig:dataset}

\end{figure}

We collected a total of 36 concepts, each of which was colored by 24 designers, resulting in 864 color samples.
The colorings are of high quality, with consistent color choices across designers.
Figure~\ref{fig:dataset} illustrates examples of the coloring results, and the corresponding extracted color compositions are shown above.
The whole dataset is available in the supplementary material, and will be released in the future.

\subsection{Quantitative Evaluation}
\label{ssec:quantitative-evaluation}
To evaluate the effectiveness of our approach, we conducted a quantitative evaluation comparing the color compositions generated by our generative method with those query-based results.
\begin{figure*}[t]
    \centering
    \includegraphics[width=0.985\linewidth]{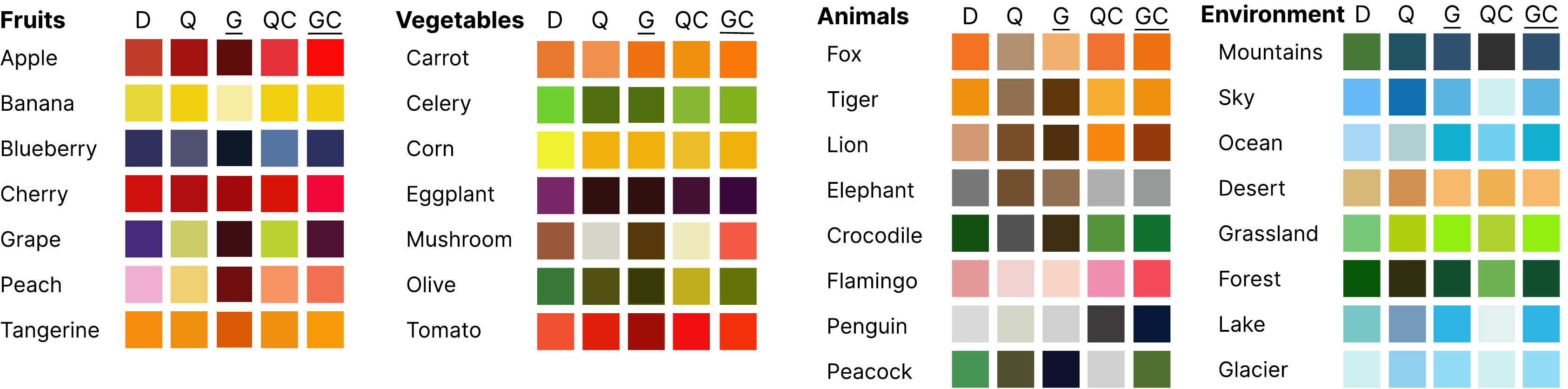}
    \caption{
        Our result compares with previous works on basic concepts.
        D stands for the designers's ground truth, G stands for the generated result by our method, Q stands for the query-based result, and QC stands for the query-based clipart result.}
    \Description{
        Our result compares with previous works on basic concepts.
        D stands for the designers's ground truth, G stands for the generated result by our method, Q stands for the query-based result, and QC stands for the query-based clipart result.
    }
    \label{fig:result_dominant}
\end{figure*}

\subsubsection{Experiment settings}
For the evaluation, we constructed a dataset of color compositions for 36 distinct concepts aligned with the designers' coloring dataset.
We included two types of image sources: photos and clipart.
Previous studies suggest that, for certain concepts, human-made illustrations may better capture people's color associations~\cite{lin2013selecting}.
For instance, the association between \q{money} and the color green is more pronounced in clipart images than in photos.
It is important to determine when to use photos and when to use clipart based on the specific concept.
For each concept, we applied the following conditions to obtain the image dataset and process them with the pipeline in Sect.~\ref{sec:gencolor}:
\begin{itemize}
    \item \textbf{Queried Image (Q)}:
          Image datasets were obtained by querying the concept in Google Image Search.
          We use the top 50 images returned from the search engine, consistent with previous studies~\cite{hu2023self}.
          The search engine is expected to return the most relevant images for the concept.
          Most of the images are photos, but some clipart images may also be included.

    \item \textbf{Generated Image (G)}:
          Realistic photo datasets generated by our method are designed to reflect the natural color patterns associated with photographic representations of the concept.
          The prompt is set to generate \q{realistic photo} with \q{natural light}.
          Contexts have been enhanced in prompt to generate more context-dependent color, such as \q{exuding a sense of depression and heaviness} for \q{polluted}.
          Note that the prompt is not directly related to the color, but the context can influence the color association.

    \item \textbf{Query-based Clipart (QC)}:
          Image dataset obtained by querying the concept with the term \q{clipart} appended in Google Image Search.
          Color compositions are generated by querying the concept with the term \q{clipart} appended in Google Image Search.
          This method filters the results to focus on more simplified and symbolic representations, emphasizing key colors typically used in clipart images.

    \item \textbf{Generated Clipart (GC)}:
          Clipart datasets generated by our method, based on simplified and stylized representations that often use more distinct and exaggerated colors.
          This dataset is generated with the prompt \q{simple flat design}. The context is also enhanced in the prompt.
\end{itemize}

\re{
    We observed that the T2I model responds more effectively to visual contexts than to descriptions targeting other senses.
    For instance, context like \q{clear} versus \q{polluted} are captured more easily than \q{lively} versus \q{quiet.} 
    This likely stems from the model's training on visual data, making it less sensitive to certain auditory terms.
    To improve this, we enhanced prompts by using more detailed descriptions related to atmosphere and emotions, such as replacing \q{quiet} with \q{evoking feelings of silence and lonely.} 
    Further details are available in the Appendix.
}

\begin{figure*}[htb]
    \centering
    \includegraphics[width=0.985\linewidth]{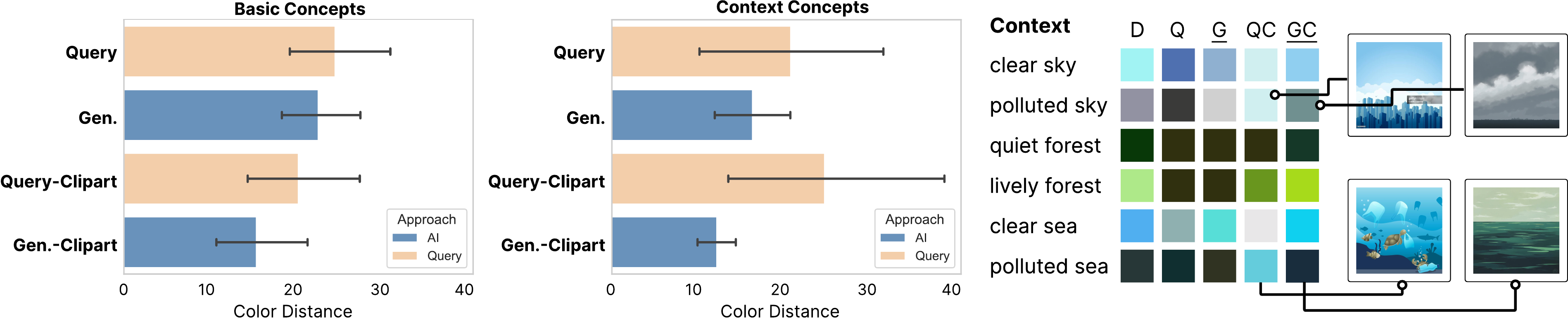}
    \caption{Color distance between different approaches and the designer's ground truth. Left shows the color distance of our generative approach and two query-based methods, across basic concepts and context-dependent concepts. Right shows the dominant color of each context, with example generated and queried images.}
    \Description{Color distance between different approaches and the designer's ground truth. Left shows the color distance of our generative approach and two query-based methods, across basic concepts and context-dependent concepts. Right shows the dominant color of each context, with example generated and queried images.}
    \label{fig:color_distance}
\end{figure*}

\subsubsection{Color Difference Comparison}
\label{sssec:color-differenece-comparison}
To demonstrate the effectiveness of our approach, we compared the color generated by our method with those obtained from the query-based methods.
First, we visually examined the primary colors of each basic concept, as shown in Figure~\ref{fig:result_dominant}.
Then, we used CIEDE2000 color difference to quantify the similarity between the primary colors extracted from the generated images and the designers' ground truth.

The results in Figure~\ref{fig:result_dominant} show that the primary colors extracted from our generated approach are generally consistent with those produced by designers.
While there are minor visual differences in brightness and saturation, the color differences remain within an acceptable range, and the obtained colors are reasonable.
In contrast, the query-based approach exhibits more variation, particularly in environmental concepts.

These observations are further supported by the quantitative color distance comparison in Figure~\ref{fig:color_distance}.
We grouped the comparisons based on the image source type—whether generated from photos or clipart—since it's not reasonable to directly compare colors between photos and clipart.
For basic concepts, our generated approach outperformed the query-based approach and aligned more closely with the designers' results.
Specifically, the generated clipart ($M=15.18$, $SD=14.90$) had a smaller color distance than the queried clipart ($M=19.97$, $SD=18.92$).
For photos, the generated image ($M=22.26$, $SD=13.03$) was also closer to the ground truth compared to the general queried results ($M=24.19$, $SD=17.61$).

When examining context-dependent concepts, the advantages of the generative approach are even more pronounced, with the small color distance and more stable results, as shown in Figure~\ref{fig:color_distance}.
The generated clipart ($M=13.37$, $SD=3.26$) had a much smaller and more stable color distance compared to the query-based clipart ($M=27.08$, $SD=18.56$).
Similarly, the generated image ($M=17.91$, $SD=6.19$) was closer to the ground truth compared to the general queried results ($M=22.77$, $SD=15.12$).
This advantage is likely due to the query-based approach's failure to properly link colors with context.
For instance, as shown in Figure~\ref{fig:color_distance} (right), when querying for clipart of a polluted sky, the results often add visual elements such as chimneys and smoke to convey pollution but make little effort to modify the actual color of the sky.
Similarly, in the clipart of a polluted sea, garbage is added to the image, but the sea remains blue.
In contrast, the colors generated by our approach are closely aligned with the context, effectively conveying the intended atmospheric or environmental feel, such as the somber tone of a polluted sky or the murky green of a polluted sea.

\subsubsection{User Rating}
\label{sssec:user-rating}
\begin{figure}[htb]
    \centering
    \includegraphics[width=0.985\linewidth]{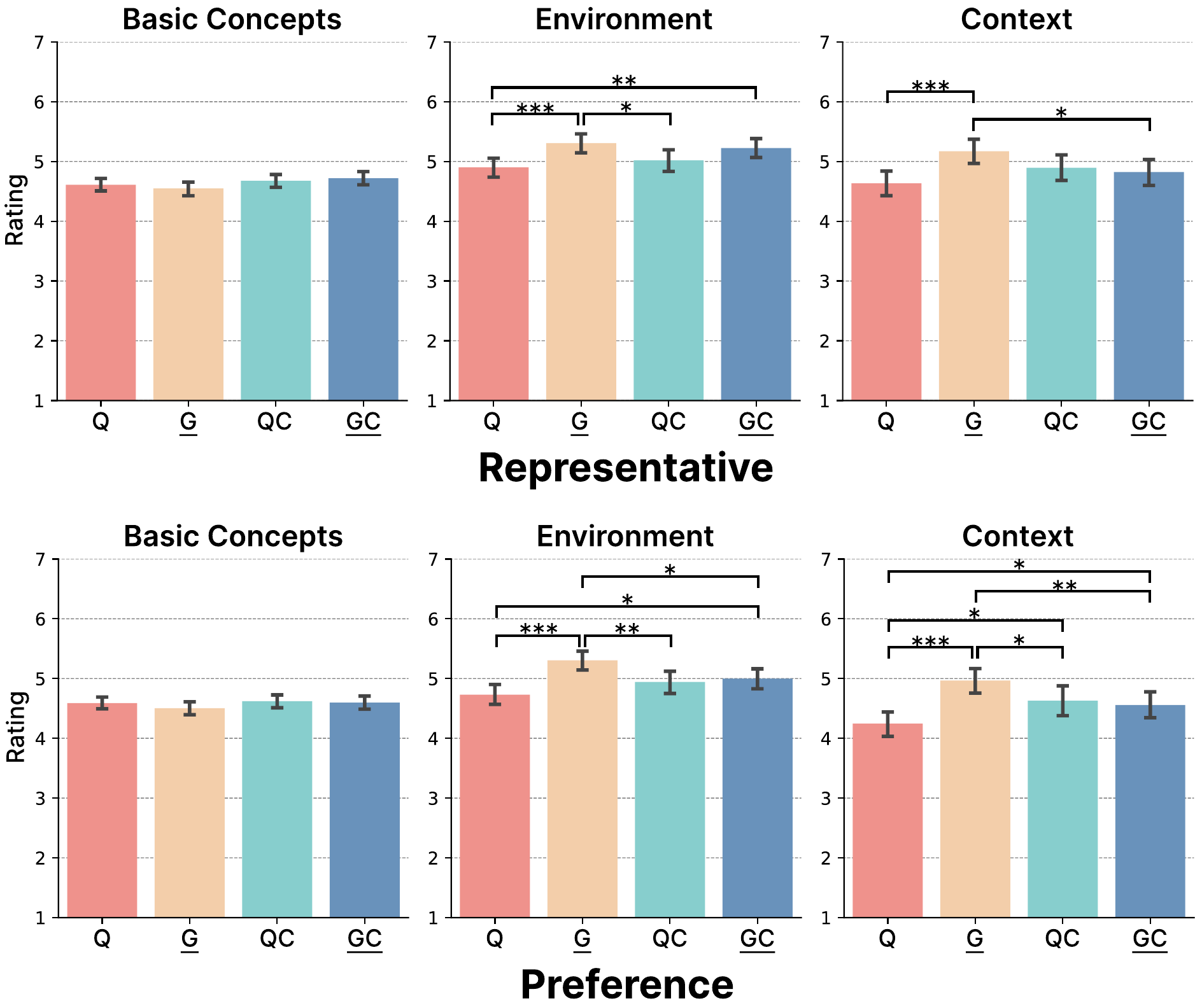}
    \caption{User rating for the representative and preference of the color-context association across four different methods and four groups of concepts.}
    \Description{User rating for the representative and preference of the color-context association across four different methods and four groups of concepts.}
    \label{fig:user_study_result}
\end{figure}

We conducted a user study to evaluate the representative and preference of the color-context association across four different methods and four groups of concepts.
The results are shown in Figure~\ref{fig:user_study_result}.

\textbf{Study Design.}
We recruited 30 participants (21 females, 9 males) through posters and social media, with an average age of 27.07 years ($SD = 6.90$).
Among the participants, 12 participants have design backgrounds, including graphic design, UI design, visual design, etc.
The remaining 18 participants came from various fields, such as engineering, computer science, and psychology, with no professional design experience.
This participant pool provided a wide range of perspectives for the evaluation.
The experiment assessed two primary factors: \textit{representativeness} and \textit{preference}, both rated on a 7-point Likert scale.
Representativeness was defined as the extent to which the color composition was perceived to align with the concept and context, while preference referred to participants' subjective liking of the color composition, both are primarily influenced by color harmony and visual appeal.
Participants evaluated color compositions generated by four different methods: Query (Q), Generated Image (G), Query Clipart (QC), and Generated Clipart (GC).
A total of 36 concepts were tested across these four methods and two metrics, resulting in 8640 trials (36 concepts $\times$ 4 methods $\times$ 2 metrics $\times$ 30 participants).
The order of the trials was fully randomized, and participants were blinded to the method used to generate each color composition.
The test took approximately 20 minutes to complete, and participants were compensated approximately 3 USD for their time.

\textbf{Results.}
We conducted a one-way ANOVA to compare the ratings across the four methods and four groups of concepts, followed by a post-hoc T-test to identify significant differences between these methods.

For the representativeness evaluation, the generated clipart received the highest rating ($M = 4.71$, $SD = 1.46$), while the generated image ($M = 4.55$, $SD = 1.47$) had the lowest rating for basic concepts. The differences between these ratings and those of the query clipart ($M = 4.68$, $SD = 1.41$) and query ($M = 4.61$, $SD = 1.37$) were minimal.
In the environmental concepts group, the generated image ($M = 5.31$, $SD = 1.26$) received the highest rating, significantly higher than both the query ($M = 4.90$, $SD = 1.27$, $p < .001$) and the query clipart ($M = 5.02$, $SD = 1.41$, $p < .05$). Additionally, the generated clipart ($M = 5.23$, $SD = 1.23$) was rated significantly higher than the query ($p < .01$).
Similarly, in the context-dependent concepts group, the generated image ($M = 5.17$, $SD = 1.44$) was rated significantly higher than both the query ($M = 4.63$, $SD = 1.41$, $p < .001$) and the generated clipart ($M = 4.82$, $SD = 1.43$, $p < .05$). Although the query clipart ($M = 4.89$, $SD = 1.50$) received a higher score than the generated clipart, the difference was not statistically significant.
These findings suggest that participants perceive the generated images as more representative than other methods, particularly for natural environments and context-dependent concepts.

\begin{figure*}[htb]
    \centering
    \includegraphics[width=0.985\linewidth]{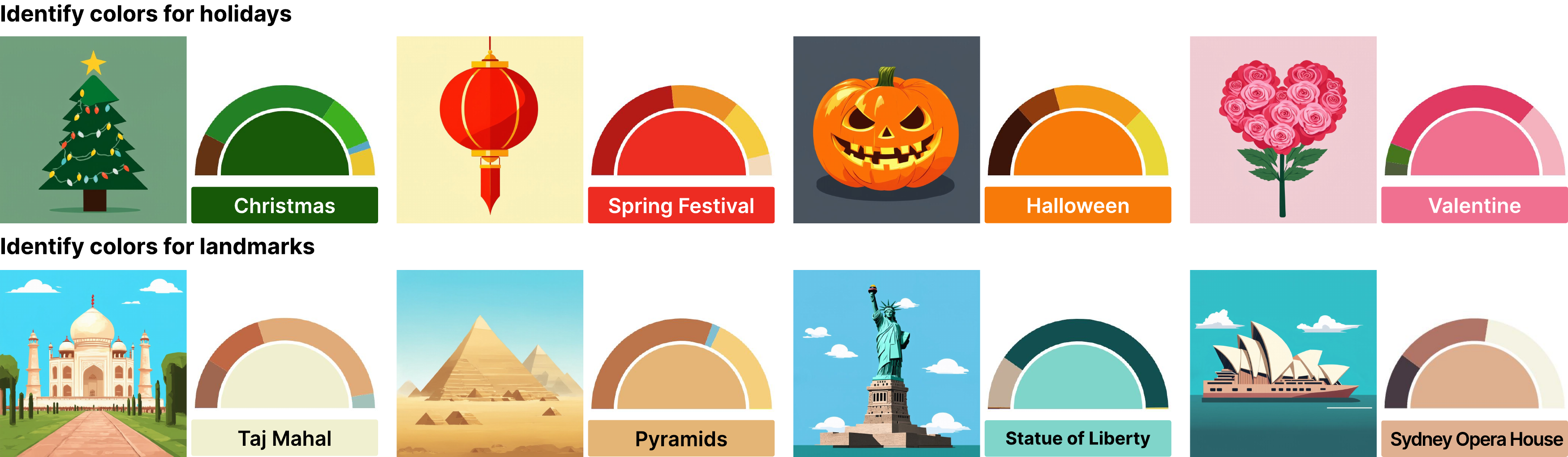}
    \caption{Example of identifying the representative color of holiday and landmark.}
    \Description{Example of identifying the representative color of holiday and landmark.}
    \label{fig:application_identify_color}
\end{figure*}

In the preference evaluation, no significant differences were found between the four approaches in the basic concepts group.
However, the query clipart approach ($M = 4.62$, $SD = 1.41$) received the highest preference rating, followed closely by the generated clipart ($M = 4.60$, $SD = 1.47$), query ($M = 4.59$, $SD = 1.33$), and generated image ($M = 4.50$, $SD = 1.41$).
For the environmental concepts group, participants significantly preferred the generated image ($M = 5.30$, $SD = 1.27$) over all other methods.
This difference was statistically significant when compared to the generated clipart ($M = 5.00$, $SD = 1.37$, $p < .05$), query ($M = 4.73$, $SD = 1.30$, $p < .001$), and query clipart ($M = 4.94$, $SD = 1.42$, $p < .01$).
In the context-dependent concepts group, the generated image again received the highest preference rating ($M = 4.97$, $SD = 1.42$), with significant differences observed compared to the query clipart ($M = 4.63$, $SD = 1.61$, $p < .05$), generated clipart ($M = 4.56$, $SD = 1.51$, $p < .01$), and query methods ($M = 4.24$, $SD = 1.43$, $p < .001$).
These findings suggest that participants prefer the generated image for natural environments and context-dependent concepts.
The generated clipart received the second-highest rating in the environmental concepts group, significantly outperforming the query method ($p < .05$).
In contrast, the query method was rated lowest in the context-dependent concepts group, with statistically significant differences compared to all other methods: generated image ($p < .001$), generated clipart ($p < .05$), and query clipart ($p < .05$).
This indicates that participants tend to dislike the query method the most for context-dependent concepts.

\begin{figure*}[htb]
    \centering
    \includegraphics[width=0.8\linewidth]{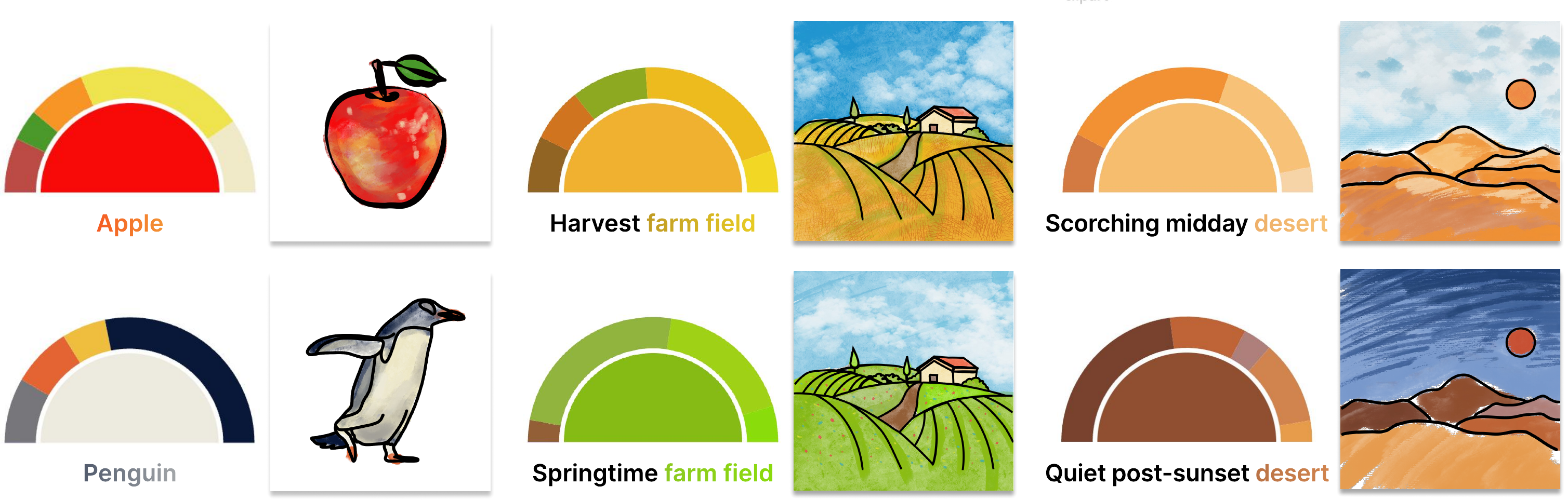}
    \caption{Result of user coloring task, demonstrating how the recommended colors are applied to different conceptual entities and contexts, such as seasonal or time variations.}
    \Description{Result of user coloring task, demonstrating how the recommended colors are applied to different conceptual entities and contexts, such as seasonal or time variations.}
    \label{fig:application_coloring}
\end{figure*}

Overall, no significant differences were observed in representativeness or preference among the four methods for basic concepts.
However, for natural environments and context-dependent concepts, the generated image outperformed the other methods in both representativeness and user preference, while the query method was rated as the least representative and least preferred.

\subsection{Application Scenario}
\label{ssec:appication-scenario}
We demonstrate two application scenarios where our method can support designers in color-concept association tasks: identifying representative colors for design elements and clipart coloring.

\subsubsection{Identifying Representative Colors for Design Elements}
In scenarios such as designing posters for holiday promotions or city branding campaigns, identifying the primary colors associated with each design element is a critical first step.
Our method offers an efficient solution for designers to pinpoint representative colors that align with the intended theme.
Figure~\ref{fig:application_identify_color} illustrates examples of identifying the representative colors for both holidays and landmarks.
For instance, considering key elements of Christmas, such as the Christmas tree, the method recommends a color composition with green as the primary color, which is strongly associated with the holiday.
Similarly, when branding iconic landmarks like the Taj Mahal or the Statue of Liberty, the method generates distinct color compositions that reflect each landmark's unique visual identity—beige and gold for the Pyramids or teal for the Statue of Liberty.
This ensures consistency and visual cohesion across a series of promotional materials.

\subsubsection{Clipart Coloring}
In another scenario, designers tasked with coloring illustrations for promotional graphics can benefit from the color compositions generated by our method. These results serve as a valuable reference, aligning closely with public perception and providing guidance for accurate and effective coloring.
Figure~\ref{fig:application_coloring} illustrates how designers apply the recommended colors to various conceptual entities and contexts.
For basic design elements, such as apples or penguins, the method offers intuitive and recognizable color compositions that enhance the clarity and visual appeal of the design.
In more complex scenarios, the method adapts its color recommendations based on environmental or temporal contexts.
For instance, a farm field during harvest is characterized by warm browns and yellows, while the same field in spring might feature fresh greens.
Similarly, a desert scene at midday could be depicted in harsh, hot, and bright tones, whereas a desert at sunset might be represented with cooler, darker tones.
This flexibility in adapting to diverse contexts not only streamlines the design process but also provides designers with a reliable starting point or reference for their work.
By offering context-dependent color recommendations, our method helps designers create more engaging and visually coherent designs that resonate with their intended audience.

\subsection{Gallery of Color-Concept Association}
\label{ssec:gallery}
\begin{figure}[htb]
    \centering
    \includegraphics[width=0.985\linewidth]{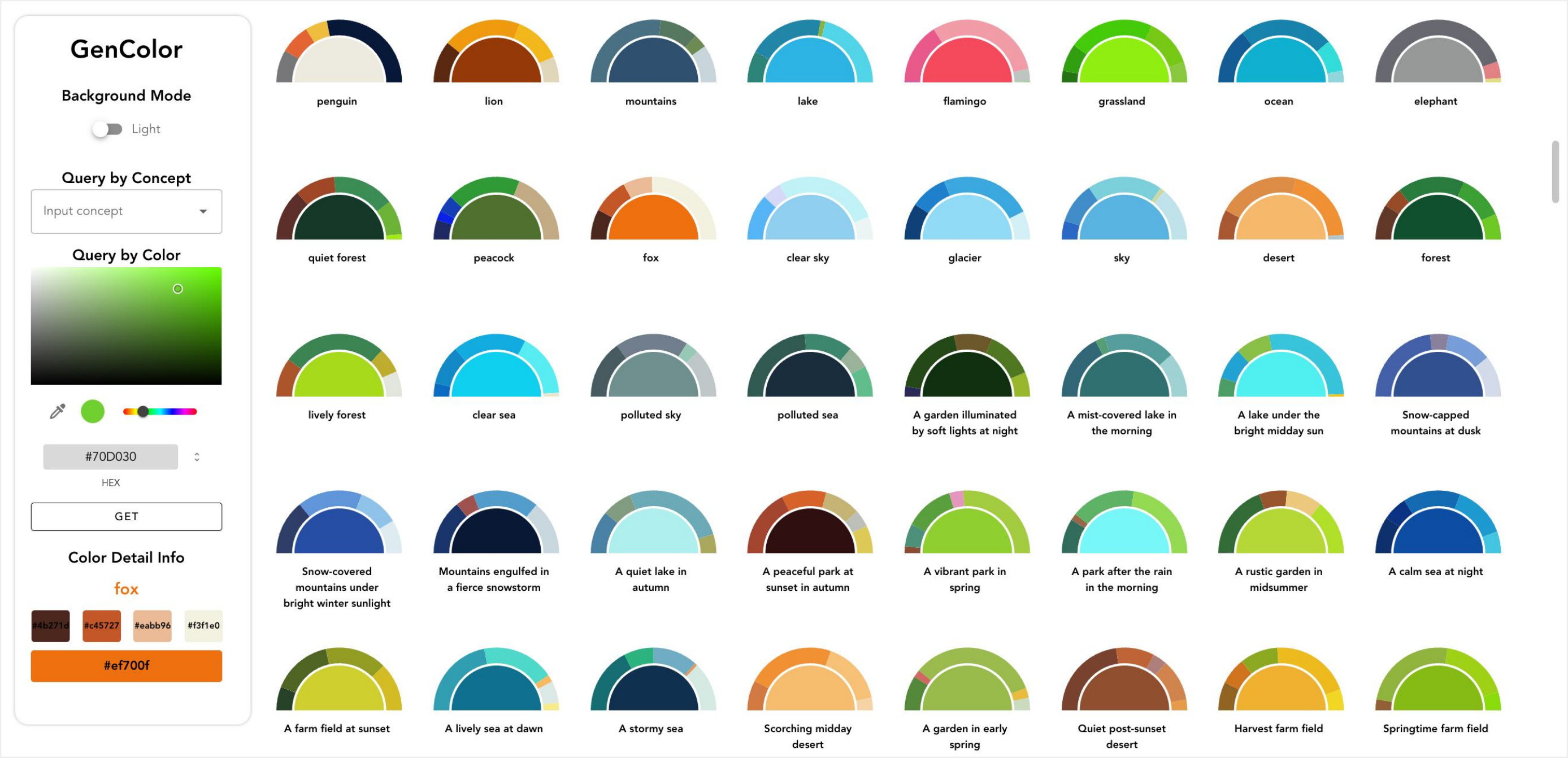}
    \caption{Gallery of color-concept associations, showcasing a wide range of concepts and contexts.}
    \Description{Gallery of color-concept associations, showcasing a wide range of concepts and contexts.}
    \label{fig:gallery}
\end{figure}

To enhance research and practical design applications, we developed a gallery of color-concept associations, offering a diverse range of concepts and contexts (see Figure~\ref{fig:gallery}).
This aligns with previous studies~\cite{kim2020lexichrome,xkcd2010color}, providing a visual representation showing the outcome of studies on color-concept associations.
Our gallery includes common design elements and extends to nuanced categories such as different styles, emotions, and times of day.
Designers can perform text-based searches to easily explore and retrieve relevant color compositions, enabling them to find colors that best suit their specific design needs.
This functionality provides designers with an extensive and adaptable resource to support various creative projects.

\section{Discussion}
\subsection{Personalized Color-concept Association}
\begin{figure}[htb]
    \centering
    \includegraphics[width=0.985\linewidth]{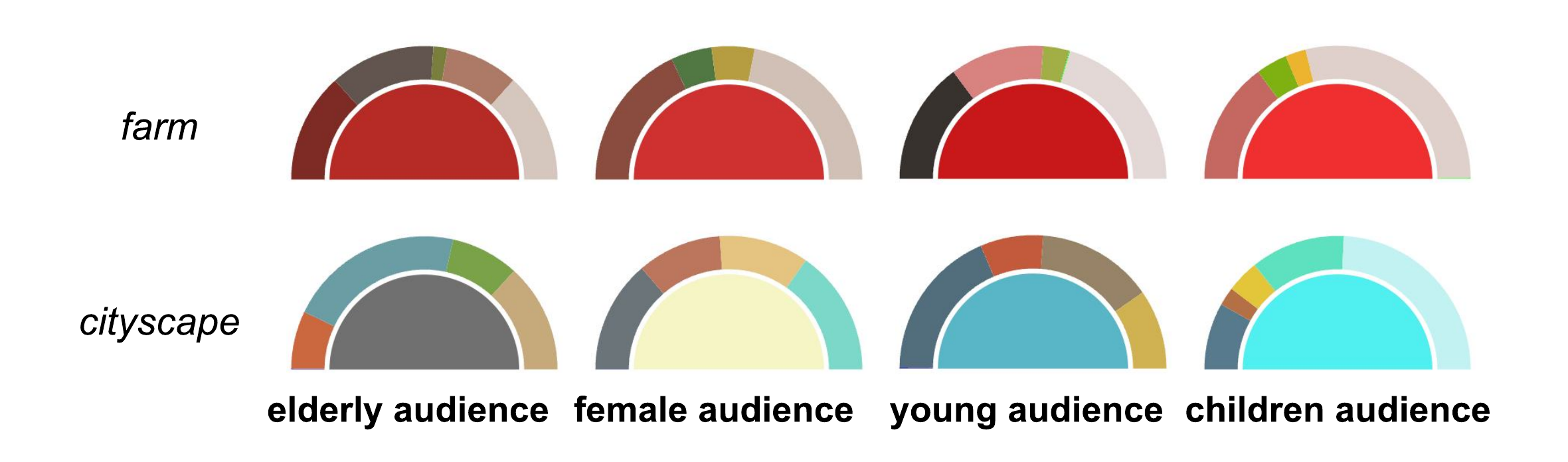}
    \caption{Associated colors for farm and cityscape concepts as perceived by different audience groups.}
    \Description{Associated colors for farm and cityscape concepts as perceived by different audience groups.}
    \label{fig:audience_group}
\end{figure}

Generative AI's ability to produce universal color-concept associations that align closely with human judgments demonstrates its reliability.
It is valuable for designers to effectively convey intended meanings and emotions to the general public, as well as to ensure that their designs are universally understood.
However, generative AI can be extended beyond universal associations to cater to more personalized design solutions.
Acting as an adaptive agent, generative AI can simulate more nuanced
audience perspectives by factoring in personality traits, cultural contexts, and emotional responses, which are essential for achieving personalized design solutions~\cite{hegemann2024palette}.
This adaptability is particularly impactful when designing for diverse audiences where a one-size-fits-all approach to color selection is insufficient.
We initially tested how different audiences, such as elderly individuals, women, young adults, and children, perceive colors in the contexts of farms and cityscapes, as shown in Figure~\ref{fig:audience_group}.
The results reveal that color related to children is bright and saturated, while for elderly audiences, it is darker and muted.
These findings highlight the capability of generative AI to tailor color associations in ways that resonate more deeply with specific user groups.

These findings highlight generative AI's potential to transcend traditional design practices, offering nuanced color associations for personalized, context-sensitive designs that align with diverse audiences' needs.

\begin{figure*}[htb]
    \centering
    \includegraphics[width=0.99\linewidth]{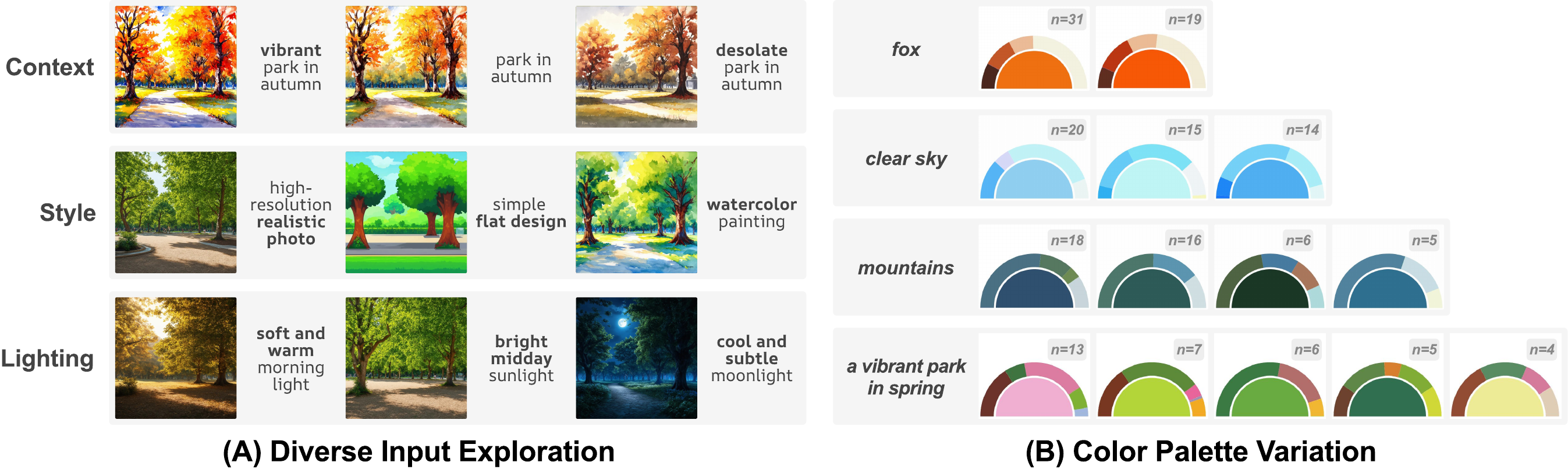}
    \caption{\re{Exploring diverse color-concept associations using GenColor.
            (A) Prompt level: GenColor allows diverse input exploration, allowing designers to specify varying attributes.
            (B) Palette level: GenColor provides a spectrum of palettes linked to each concept, offering designers a broad range of options.}}
    \Description{Exploring diverse color-concept associations using GenColor.
        (A) Prompt level: facilitates diverse input exploration, allowing designers to specify varying attributes.
        (B) Palette level: provides a spectrum of palettes linked to each concept, offering designers a broad range of options.}
    \label{fig:diversity-case}
\end{figure*}

\subsection{Balancing Creativity and Consistency}
The integration of generative AI into design workflows reveals a critical tension between creativity and consistency.
On the one hand, AI's capacity to generate consistent color-concept associations ensures designs are broadly accessible and effectively communicate intended messages, aligning with general audience expectations.
This is particularly beneficial for conveying widely accepted meanings or evoking predictable emotional responses, enhancing efficiency in reaching diverse audiences.
While consistency ensures baseline communicability, controlled diversity at the palette level allows creative exploration within defined boundaries.
Design is not solely about meeting audience expectations; it also involves innovation and reflecting the designer's unique perspective~\cite{nelson2014design}.
Thus, it is essential that generative AI supports diverse color-concept associations without hindering designers' creativity.

By introducing diversity at the prompt and color palette levels, GenColor facilitates the exploration of diverse color-concept associations.
At the prompt level, we enhance creative flexibility through a variety of context-aware concepts (\eg, vibrant and desolated), styles (\eg, photo, flat design, painting), and lighting (\eg, tones and intensity) options, as shown in Figure~\ref{fig:diversity-case} (A).
Different prompts can influence image generation, capturing diverse design intentions even with a fixed seed.
We acknowledge that a concept may be linked to multiple color palettes for color variation at the palette level.
As illustrated in Figure~\ref{fig:diversity-case} (B), we provide a spectrum of palettes associated with each concept, giving designers a wide array of options.
Concrete concepts like \q{fox} or \q{clear sky} may show slight shade variations. In contrast, broader concepts like \q{mountains} or \q{park in spring} lead to more diverse palettes featuring a wider range of shades and tones.
For instance, mountains might include various greens and blues hues, while spring parks could showcase pinks, greens, and yellows.

Moreover, generated color palettes are intended to support designers' decisions by serving as inspirational references, not as final solutions.
Importantly, designers should always retain the freedom to select, modify, and refine these suggestions according to their vision.
Generative AI should be seen as offering a starting point or reference rather than as a substitute for the designer's voice.
In this collaborative process, the role of the designer is crucial in refining and personalizing AI suggestions, ensuring that the final design embodies both consistency and creativity.
In this way, AI can facilitate the efficient generation of typical color associations while still allowing designers to express their unique creative identity.

\subsection{Limitation and Future Work}

\textbf{Expand the Scope of Concepts.}
Our current framework primarily addresses concrete concepts within specific contexts. However, abstract concepts, such as \q{danger} or \q{freedom} are challenging to represent visually and remain outside the scope of this study. To improve, incorporating metaphorical or symbolic imagery could better capture these abstract ideas. Additionally, our approach currently handles single concepts in isolation and does not address the complexities of managing multiple concepts simultaneously. In the formative study, E4 highlighted the difficulty of selecting colors for multiple concepts while maintaining harmony and conveying the intended context effectively. Future work could focus on methods to support the composition of colors representing multiple concepts, ensuring cohesive and harmonious design outcomes.

\textbf{Enhance Contextual Adaptability.}
Our current testing has focused on conditional and emotional contexts.
While our framework is theoretically generalizable to more complex contexts—such as audience preferences, cultural influences, or artistic styles—these areas have not been sufficiently explored.
Future work could investigate how generative AI can adapt to these richer contexts and offer relevant color recommendations.
An intriguing area for future research is to explore which contexts are interpretable by generative AI. Understanding this could provide valuable insights into optimizing AI for diverse and nuanced design needs.

\textbf{Biases and Limitations in T2I Model.}
Despite generating high-quality images, T2I models exhibit North-centric biases, where AI systems are disproportionately influenced by data and perspectives from developed regions~\cite{naik2023social,wan2024survey,prerak2024addressing}.
This bias can also affect color-concept associations.
For instance, the color generated for \q{money} is typically green, reflecting its association with the US dollar, and \q{taxis} are generally portrayed as yellow.
However, as artificial concepts, these colors can vary across different countries.
In our approach, we address these variations by associating each concept with multiple color palettes, allowing the model to capture a broader range of color associations.
In future work, thoroughly exploring model biases from the perspective of color is essential.
Another limitation of the T2I model is its sensitivity to different prompts.  While we have made efforts to enhance the model's responsiveness by refining prompt structures and incorporating more detailed context, as mentioned in Sect.~\ref{ssec:quantitative-evaluation}, the model still doesn't perform perfectly across all concepts.
Therefore, it is essential to recognize that generative AI is a supportive tool that assists designers by offering suggestions, rather than replacing their expertise.

\textbf{Expanding and Comparing Generative Models.}
This work focuses on \re{T2I} models for generating images to extract color-concept associations, which are preferred by designers as they provide direct references for color selection and align with design workflows.
Additionally, large language models also show the capability to generate color-concept associations based on text descriptions.
Previous research has demonstrated that color-concept associations derived from language models generally correlate with human judgments~\cite{mukherjee2024large}.
Exploring the differences between image-based and language-based references is valuable. Future work should compare these approaches to gain insights into their interpretability and investigate the potential benefits of combining them.

\textbf{Ethical Considerations of GenAI Tools.}
\rec{While GenColor aims to support designers in color exploration, we acknowledge the broader ethical implications of GenAI tools.
    One concern is GenAI tools may lead to the homogenization of creative output, producing works that conform to mainstream designers and diminishing diversity~\cite{oppenlaender2022creativity,samuelson2023generative,gautam2024melting}.
    Marginalized designers, whose works often reflect unique personal and cultural elements, may be disproportionately affected.
    Additionally, the automation of some tasks traditionally performed by human designers could lead to job losses and exacerbate inequality in the design industry, especially for those marginalized ones~\cite{samuelson2023generative}.
    The HCI community has recognized that GenAI tools should serve as supportive rather than replacement tools for human creativity~\cite{oppenlaender2022creativity}.
    They advocate for keeping designers in the loop, enabling them to leverage GenAI tools to enhance efficiency in routine tasks and thus devote more time and energy to creative and expressive aspects of their work~\cite{hwang2022late,lee2024when}.
    Aligning with these principles, GenColor is positioned as an assistive tool that empowers designers rather than replacing their creative role.
    Future work should focus on fostering diversity, ensuring transparency, and establishing ethical guidelines to promote responsible and inclusive use of GenAI in design practices.
}

\section{Conclusion}

In this study, we introduced \textit{GenColor}, a generative framework designed to improve the association of colors with concepts in visual design. 
By leveraging \re{T2I} models, \textit{GenColor} overcomes the limitations of traditional retrieval-based methods, offering flexibility in generating context-dependent images and robustness in managing variations.
Our evaluation shows that \textit{GenColor} aligns closely with designer preferences and outperforms existing approaches. 
We also provide a publicly available gallery of color-concept associations to support research and practical applications in design.
Overall, \textit{GenColor} enhances designers' ability to make informed and contextually relevant color choices, facilitating more effective visual communication.


\begin{acks}
  The authors wish to thank the anonymous reviewers for their valuable comments, as well as the time and effort of all the study participants.
  This work is partially supported by the Guangzhou Basic and Applied Basic Research Foundation (No. 2024A04J6462).
\end{acks}

\bibliographystyle{ACM-Reference-Format}
\bibliography{sample-base}


\begin{thebibliography}{81}


\ifx \showCODEN    \undefined \def \showCODEN     #1{\unskip}     \fi
\ifx \showISBNx    \undefined \def \showISBNx     #1{\unskip}     \fi
\ifx \showISBNxiii \undefined \def \showISBNxiii  #1{\unskip}     \fi
\ifx \showISSN     \undefined \def \showISSN      #1{\unskip}     \fi
\ifx \showLCCN     \undefined \def \showLCCN      #1{\unskip}     \fi
\ifx \shownote     \undefined \def \shownote      #1{#1}          \fi
\ifx \showarticletitle \undefined \def \showarticletitle #1{#1}   \fi
\ifx \showURL      \undefined \def \showURL       {\relax}        \fi
\providecommand\bibfield[2]{#2}
\providecommand\bibinfo[2]{#2}
\providecommand\natexlab[1]{#1}
\providecommand\showeprint[2][]{arXiv:#2}

\bibitem[Adobe({[n.\,d.]})]%
        {adobeSpectrumColors}
\bibfield{author}{\bibinfo{person}{Adobe}.}
  \bibinfo{year}{[n.\,d.]}\natexlab{}.
\newblock \bibinfo{title}{Reinventing Adobe Spectrum's colors}.
\newblock
  \bibinfo{howpublished}{\url{https://adobe.design/stories/design-for-scale/reinventing-adobe-spectrum-s-colors}}.
\newblock
\newblock
\shownote{Accessed: 2024-12-01}.


\bibitem[Azizi et~al\mbox{.}(2023)]%
        {azizi2023synthetic}
\bibfield{author}{\bibinfo{person}{Shekoofeh Azizi}, \bibinfo{person}{Simon
  Kornblith}, \bibinfo{person}{Chitwan Saharia}, \bibinfo{person}{Mohammad
  Norouzi}, {and} \bibinfo{person}{David~J Fleet}.}
  \bibinfo{year}{2023}\natexlab{}.
\newblock \bibinfo{title}{Synthetic Data from Diffusion Models Improves
  ImageNet Classification}.
\newblock


\bibitem[Ban et~al\mbox{.}(2024)]%
        {ban2024understanding}
\bibfield{author}{\bibinfo{person}{Yuanhao Ban}, \bibinfo{person}{Ruochen
  Wang}, \bibinfo{person}{Tianyi Zhou}, \bibinfo{person}{Minhao Cheng},
  \bibinfo{person}{Boqing Gong}, {and} \bibinfo{person}{Cho-Jui Hsieh}.}
  \bibinfo{year}{2024}\natexlab{}.
\newblock \showarticletitle{Understanding the Impact of Negative Prompts: When
  and How Do They Take Effect?}
\newblock \bibinfo{journal}{\emph{arXiv preprint arXiv:2406.02965}}
  (\bibinfo{year}{2024}).
\newblock


\bibitem[Berlin and Kay(1991)]%
        {berlin1991basic}
\bibfield{author}{\bibinfo{person}{Brent Berlin} {and} \bibinfo{person}{Paul
  Kay}.} \bibinfo{year}{1991}\natexlab{}.
\newblock \bibinfo{booktitle}{\emph{Basic Color Terms: Their Universality and
  Evolution}}.
\newblock \bibinfo{publisher}{Univ of California Press}.
\newblock


\bibitem[Brooks et~al\mbox{.}(2023)]%
        {brooks2023instructpix2pix}
\bibfield{author}{\bibinfo{person}{Tim Brooks}, \bibinfo{person}{Aleksander
  Holynski}, {and} \bibinfo{person}{Alexei~A Efros}.}
  \bibinfo{year}{2023}\natexlab{}.
\newblock \showarticletitle{{InstructPix2Pix}: Learning to Follow Image Editing
  Instructions}. In \bibinfo{booktitle}{\emph{Proceedings of the IEEE/CVF
  conference on computer vision and pattern recognition}}.
  \bibinfo{pages}{18392--18402}.
\newblock
\href{https://doi.org/10.1109/CVPR52729.2023.01764}{doi:\nolinkurl{10.1109/CVPR52729.2023.01764}}


\bibitem[Chang et~al\mbox{.}(2015)]%
        {chang2015palette}
\bibfield{author}{\bibinfo{person}{Huiwen Chang}, \bibinfo{person}{Ohad Fried},
  \bibinfo{person}{Yiming Liu}, \bibinfo{person}{Stephen DiVerdi}, {and}
  \bibinfo{person}{Adam Finkelstein}.} \bibinfo{year}{2015}\natexlab{}.
\newblock \showarticletitle{Palette-based Photo Recoloring.}
\newblock \bibinfo{journal}{\emph{ACM Trans. Graph.}} \bibinfo{volume}{34},
  \bibinfo{number}{4} (\bibinfo{year}{2015}), \bibinfo{pages}{139--1}.
\newblock
\href{https://doi.org/10.1145/2766978}{doi:\nolinkurl{10.1145/2766978}}


\bibitem[Chen et~al\mbox{.}(2023)]%
        {chen2023s}
\bibfield{author}{\bibinfo{person}{Eric~Ming Chen}, \bibinfo{person}{Jin Sun},
  \bibinfo{person}{Apoorv Khandelwal}, \bibinfo{person}{Dani Lischinski},
  \bibinfo{person}{Noah Snavely}, {and} \bibinfo{person}{Hadar Averbuch-Elor}.}
  \bibinfo{year}{2023}\natexlab{}.
\newblock \showarticletitle{What's in a Decade? Transforming Faces Through
  Time}. In \bibinfo{booktitle}{\emph{Computer Graphics Forum}},
  Vol.~\bibinfo{volume}{42}. \bibinfo{pages}{281--291}.
\newblock
\href{https://doi.org/10.1111/cgf.14761}{doi:\nolinkurl{10.1111/cgf.14761}}


\bibitem[Chen et~al\mbox{.}(2024)]%
        {chen2024oppurtunities}
\bibfield{author}{\bibinfo{person}{Minshuo Chen}, \bibinfo{person}{Song Mei},
  \bibinfo{person}{Jianqing Fan}, {and} \bibinfo{person}{Mengdi Wang}.}
  \bibinfo{year}{2024}\natexlab{}.
\newblock \showarticletitle{Opportunities and challenges of diffusion models
  for generative AI}.
\newblock \bibinfo{journal}{\emph{National Science Review}}
  \bibinfo{volume}{11}, \bibinfo{number}{12} (\bibinfo{date}{10}
  \bibinfo{year}{2024}), \bibinfo{pages}{nwae348}.
\newblock
\showISSN{2095-5138}
\href{https://doi.org/10.1093/nsr/nwae348}{doi:\nolinkurl{10.1093/nsr/nwae348}}
\showeprint{https://academic.oup.com/nsr/article-pdf/11/12/nwae348/60677321/nwae348.pdf}


\bibitem[Chen et~al\mbox{.}(2021)]%
        {chen2021investigation}
\bibfield{author}{\bibinfo{person}{Yun Chen}, \bibinfo{person}{Luwen Yu},
  \bibinfo{person}{Stephen Westland}, {and} \bibinfo{person}{Vien Cheung}.}
  \bibinfo{year}{2021}\natexlab{}.
\newblock \showarticletitle{Investigation of Designers' Colour Selection
  Process}.
\newblock \bibinfo{journal}{\emph{Color Research \& Application}}
  \bibinfo{volume}{46}, \bibinfo{number}{3} (\bibinfo{year}{2021}),
  \bibinfo{pages}{557--565}.
\newblock
\href{https://doi.org/10.1002/col.22631}{doi:\nolinkurl{10.1002/col.22631}}


\bibitem[Clark et~al\mbox{.}(2013)]%
        {clark2013statistical}
\bibfield{author}{\bibinfo{person}{Alexander Clark}, \bibinfo{person}{Gianluca
  Giorgolo}, {and} \bibinfo{person}{Shalom Lappin}.}
  \bibinfo{year}{2013}\natexlab{}.
\newblock \showarticletitle{Statistical Representation of Grammaticality
  Judgements: the Limits of N-Gram Models}. In
  \bibinfo{booktitle}{\emph{Proceedings of the Annual Workshop on Cognitive
  Modeling and Computational Linguistics (CMCL)}}. \bibinfo{pages}{28--36}.
\newblock
\urldef\tempurl%
\url{https://aclanthology.org/W13-2604}
\showURL{%
\tempurl}


\bibitem[Dalens et~al\mbox{.}(2019)]%
        {dalens2019bilinear}
\bibfield{author}{\bibinfo{person}{Th{\'e}ophile Dalens},
  \bibinfo{person}{Mathieu Aubry}, {and} \bibinfo{person}{Josef Sivic}.}
  \bibinfo{year}{2019}\natexlab{}.
\newblock \showarticletitle{Bilinear Image Translation for Temporal Analysis of
  Photo Collections}.
\newblock \bibinfo{journal}{\emph{IEEE Transactions on Pattern Analysis and
  Machine Intelligence}} \bibinfo{volume}{43}, \bibinfo{number}{4}
  (\bibinfo{year}{2019}), \bibinfo{pages}{1197--1212}.
\newblock
\href{https://doi.org/10.1109/TPAMI.2019.2950317}{doi:\nolinkurl{10.1109/TPAMI.2019.2950317}}


\bibitem[de~l'Eclairage(1978)]%
        {commission1978recommendations}
\bibfield{author}{\bibinfo{person}{Commission~Internationale de l'Eclairage}.}
  \bibinfo{year}{1978}\natexlab{}.
\newblock \showarticletitle{Recommendations on uniform color spaces,
  color-difference equations, psychometric color therms}. CIE.
\newblock


\bibitem[Deng et~al\mbox{.}(2009)]%
        {deng2009imagenet}
\bibfield{author}{\bibinfo{person}{Jia Deng}, \bibinfo{person}{Wei Dong},
  \bibinfo{person}{Richard Socher}, \bibinfo{person}{Li-Jia Li},
  \bibinfo{person}{Kai Li}, {and} \bibinfo{person}{Li Fei-Fei}.}
  \bibinfo{year}{2009}\natexlab{}.
\newblock \showarticletitle{ImageNet: A Large-scale Hierarchical Image
  Database}. In \bibinfo{booktitle}{\emph{2009 IEEE Conference on Computer
  Vision and Pattern Recognition}}. \bibinfo{pages}{248--255}.
\newblock
\href{https://doi.org/10.1109/CVPR.2009.5206848}{doi:\nolinkurl{10.1109/CVPR.2009.5206848}}


\bibitem[Fan et~al\mbox{.}(2024)]%
        {fan2024contextcam}
\bibfield{author}{\bibinfo{person}{Xianzhe Fan}, \bibinfo{person}{Zihan Wu},
  \bibinfo{person}{Chun Yu}, \bibinfo{person}{Fenggui Rao},
  \bibinfo{person}{Weinan Shi}, {and} \bibinfo{person}{Teng Tu}.}
  \bibinfo{year}{2024}\natexlab{}.
\newblock \showarticletitle{ContextCam: Bridging Context Awareness with
  Creative Human-AI Image Co-Creation}. In
  \bibinfo{booktitle}{\emph{Proceedings of the CHI Conference on Human Factors
  in Computing Systems}}. \bibinfo{pages}{1--17}.
\newblock
\href{https://doi.org/10.1145/3613904.3642129}{doi:\nolinkurl{10.1145/3613904.3642129}}


\bibitem[Gautam et~al\mbox{.}(2024)]%
        {gautam2024melting}
\bibfield{author}{\bibinfo{person}{Sanjana Gautam},
  \bibinfo{person}{Pranav~Narayanan Venkit}, {and} \bibinfo{person}{Sourojit
  Ghosh}.} \bibinfo{year}{2024}\natexlab{}.
\newblock \showarticletitle{From melting pots to misrepresentations: Exploring
  harms in generative ai}.
\newblock \bibinfo{journal}{\emph{arXiv preprint arXiv:2403.10776}}
  (\bibinfo{year}{2024}).
\newblock


\bibitem[Goodfellow et~al\mbox{.}(2014)]%
        {goodfellow2014generative}
\bibfield{author}{\bibinfo{person}{Ian Goodfellow}, \bibinfo{person}{Jean
  Pouget-Abadie}, \bibinfo{person}{Mehdi Mirza}, \bibinfo{person}{Bing Xu},
  \bibinfo{person}{David Warde-Farley}, \bibinfo{person}{Sherjil Ozair},
  \bibinfo{person}{Aaron Courville}, {and} \bibinfo{person}{Yoshua Bengio}.}
  \bibinfo{year}{2014}\natexlab{}.
\newblock \showarticletitle{Generative Adversarial Nets}. In
  \bibinfo{booktitle}{\emph{Advances in Neural Information Processing
  Systems}}, \bibfield{editor}{\bibinfo{person}{Z.~Ghahramani},
  \bibinfo{person}{M.~Welling}, \bibinfo{person}{C.~Cortes},
  \bibinfo{person}{N.~Lawrence}, {and} \bibinfo{person}{K.Q. Weinberger}}
  (Eds.), Vol.~\bibinfo{volume}{27}. \bibinfo{publisher}{Curran Associates,
  Inc.}
\newblock
\urldef\tempurl%
\url{https://proceedings.neurips.cc/paper_files/paper/2014/file/5ca3e9b122f61f8f06494c97b1afccf3-Paper.pdf}
\showURL{%
\tempurl}


\bibitem[Hao et~al\mbox{.}(2023)]%
        {hao2024optimizing}
\bibfield{author}{\bibinfo{person}{Yaru Hao}, \bibinfo{person}{Zewen Chi},
  \bibinfo{person}{Li Dong}, {and} \bibinfo{person}{Furu Wei}.}
  \bibinfo{year}{2023}\natexlab{}.
\newblock \showarticletitle{Optimizing Prompts for Text-to-Image Generation}.
  In \bibinfo{booktitle}{\emph{Advances in Neural Information Processing
  Systems}}, Vol.~\bibinfo{volume}{36}. \bibinfo{publisher}{Curran Associates,
  Inc.}, \bibinfo{pages}{66923--66939}.
\newblock
\urldef\tempurl%
\url{https://proceedings.neurips.cc/paper_files/paper/2023/file/d346d91999074dd8d6073d4c3b13733b-Paper-Conference.pdf}
\showURL{%
\tempurl}


\bibitem[Hegemann and Oulasvirta(2024)]%
        {hegemann2024palette}
\bibfield{author}{\bibinfo{person}{Lena Hegemann} {and} \bibinfo{person}{Antti
  Oulasvirta}.} \bibinfo{year}{2024}\natexlab{}.
\newblock \showarticletitle{Palette, Purpose, Prototype: The Three Ps of Color
  Design and How Designers Navigate Them}. In
  \bibinfo{booktitle}{\emph{Proceedings of the CHI Conference on Human Factors
  in Computing Systems}}. \bibinfo{publisher}{Association for Computing
  Machinery}, Article \bibinfo{articleno}{147}, \bibinfo{numpages}{19}~pages.
\newblock
\showISBNx{9798400703300}
\href{https://doi.org/10.1145/3613904.3641976}{doi:\nolinkurl{10.1145/3613904.3641976}}


\bibitem[Hou et~al\mbox{.}(2024)]%
        {hou2024c2ideas}
\bibfield{author}{\bibinfo{person}{Yihan Hou}, \bibinfo{person}{Manling Yang},
  \bibinfo{person}{Hao Cui}, \bibinfo{person}{Lei Wang}, \bibinfo{person}{Jie
  Xu}, {and} \bibinfo{person}{Wei Zeng}.} \bibinfo{year}{2024}\natexlab{}.
\newblock \showarticletitle{C2Ideas: Supporting Creative Interior Color Design
  Ideation with a Large Language Model}. In
  \bibinfo{booktitle}{\emph{Proceedings of the CHI Conference on Human Factors
  in Computing Systems}}. \bibinfo{publisher}{Association for Computing
  Machinery}, Article \bibinfo{articleno}{172}, \bibinfo{numpages}{18}~pages.
\newblock
\showISBNx{9798400703300}
\href{https://doi.org/10.1145/3613904.3642224}{doi:\nolinkurl{10.1145/3613904.3642224}}


\bibitem[Hu et~al\mbox{.}(2023)]%
        {hu2023self}
\bibfield{author}{\bibinfo{person}{Ruizhen Hu}, \bibinfo{person}{Ziqi Ye},
  \bibinfo{person}{Bin Chen}, \bibinfo{person}{Oliver van Kaick}, {and}
  \bibinfo{person}{Hui Huang}.} \bibinfo{year}{2023}\natexlab{}.
\newblock \showarticletitle{Self-Supervised Color-Concept Association via Image
  Colorization}.
\newblock \bibinfo{journal}{\emph{IEEE Transactions on Visualization and
  Computer Graphics}} \bibinfo{volume}{29}, \bibinfo{number}{1}
  (\bibinfo{year}{2023}), \bibinfo{pages}{247--256}.
\newblock
\href{https://doi.org/10.1109/TVCG.2022.3209481}{doi:\nolinkurl{10.1109/TVCG.2022.3209481}}


\bibitem[Humphrey(2019)]%
        {humphrey2019colour}
\bibfield{author}{\bibinfo{person}{Nicholas Humphrey}.}
  \bibinfo{year}{2019}\natexlab{}.
\newblock \bibinfo{booktitle}{\emph{Colour for Architecture Today}}.
\newblock \bibinfo{publisher}{Taylor \& Francis}. 9--12 pages.
\newblock


\bibitem[Hwang(2022)]%
        {hwang2022late}
\bibfield{author}{\bibinfo{person}{Angel Hsing-Chi Hwang}.}
  \bibinfo{year}{2022}\natexlab{}.
\newblock \showarticletitle{Too Late to be Creative? AI-Empowered Tools in
  Creative Processes}. In \bibinfo{booktitle}{\emph{Extended Abstracts of the
  2022 CHI Conference on Human Factors in Computing Systems}} (New Orleans, LA,
  USA) \emph{(\bibinfo{series}{CHI EA '22})}. \bibinfo{publisher}{Association
  for Computing Machinery}, \bibinfo{address}{New York, NY, USA}, Article
  \bibinfo{articleno}{38}, \bibinfo{numpages}{9}~pages.
\newblock
\showISBNx{9781450391566}
\href{https://doi.org/10.1145/3491101.3503549}{doi:\nolinkurl{10.1145/3491101.3503549}}


\bibitem[Jahanian et~al\mbox{.}(2017)]%
        {jahanian2017colors}
\bibfield{author}{\bibinfo{person}{Ali Jahanian}, \bibinfo{person}{Shaiyan
  Keshvari}, \bibinfo{person}{S.~V.~N. Vishwanathan}, {and}
  \bibinfo{person}{Jan~P. Allebach}.} \bibinfo{year}{2017}\natexlab{}.
\newblock \showarticletitle{Colors -- Messengers of Concepts: Visual Design
  Mining for Learning Color Semantics}.
\newblock \bibinfo{journal}{\emph{ACM Trans. Comput.-Hum. Interact.}}
  \bibinfo{volume}{24}, \bibinfo{number}{1} (\bibinfo{year}{2017}).
\newblock
\showISSN{1073-0516}
\href{https://doi.org/10.1145/3009924}{doi:\nolinkurl{10.1145/3009924}}


\bibitem[Jonauskaite et~al\mbox{.}(2019)]%
        {jonauskaite2019machine}
\bibfield{author}{\bibinfo{person}{Domicele Jonauskaite},
  \bibinfo{person}{Jörg Wicker}, \bibinfo{person}{Christine Mohr},
  \bibinfo{person}{Nele Dael}, \bibinfo{person}{Jelena Havelka},
  \bibinfo{person}{Marietta Papadatou-Pastou}, \bibinfo{person}{Meng Zhang},
  {and} \bibinfo{person}{Daniel Oberfeld}.} \bibinfo{year}{2019}\natexlab{}.
\newblock \showarticletitle{A Machine Learning Approach to Quantify the
  Specificity of Colour–emotion Associations and Their Cultural Differences}.
\newblock \bibinfo{journal}{\emph{Royal Society Open Science}}
  \bibinfo{volume}{6}, \bibinfo{number}{9} (\bibinfo{year}{2019}),
  \bibinfo{pages}{190741}.
\newblock
\href{https://doi.org/10.1098/rsos.190741}{doi:\nolinkurl{10.1098/rsos.190741}}


\bibitem[Kawar et~al\mbox{.}(2023)]%
        {kawar2023imagic}
\bibfield{author}{\bibinfo{person}{Bahjat Kawar}, \bibinfo{person}{Shiran
  Zada}, \bibinfo{person}{Oran Lang}, \bibinfo{person}{Omer Tov},
  \bibinfo{person}{Huiwen Chang}, \bibinfo{person}{Tali Dekel},
  \bibinfo{person}{Inbar Mosseri}, {and} \bibinfo{person}{Michal Irani}.}
  \bibinfo{year}{2023}\natexlab{}.
\newblock \showarticletitle{Imagic: Text-Based Real Image Editing with
  Diffusion Models}. In \bibinfo{booktitle}{\emph{Proceedings of the IEEE/CVF
  Conference on Computer Vision and Pattern Recognition}}.
  \bibinfo{pages}{6007--6017}.
\newblock
\href{https://doi.org/10.1109/CVPR52729.2023.00582}{doi:\nolinkurl{10.1109/CVPR52729.2023.00582}}


\bibitem[Kim et~al\mbox{.}(2020)]%
        {kim2020lexichrome}
\bibfield{author}{\bibinfo{person}{Chris Kim}, \bibinfo{person}{Uta Hinrichs},
  \bibinfo{person}{Saif~M. Mohammad}, {and} \bibinfo{person}{Christopher
  Collins}.} \bibinfo{year}{2020}\natexlab{}.
\newblock \showarticletitle{Lexichrome: Text Construction and Lexical Discovery
  with Word-Color Associations Using Interactive Visualization}. In
  \bibinfo{booktitle}{\emph{Proceedings of the ACM Designing Interactive
  Systems Conference}} (Eindhoven, Netherlands) \emph{(\bibinfo{series}{DIS
  '20})}. \bibinfo{publisher}{Association for Computing Machinery},
  \bibinfo{address}{New York, NY, USA}, \bibinfo{pages}{477–488}.
\newblock
\showISBNx{9781450369749}
\href{https://doi.org/10.1145/3357236.3395503}{doi:\nolinkurl{10.1145/3357236.3395503}}


\bibitem[Kim et~al\mbox{.}(2014)]%
        {kim2014perceptually}
\bibfield{author}{\bibinfo{person}{Hye-Rin Kim}, \bibinfo{person}{Min-Joon
  Yoo}, \bibinfo{person}{Henry Kang}, {and} \bibinfo{person}{In-Kwon Lee}.}
  \bibinfo{year}{2014}\natexlab{}.
\newblock \showarticletitle{Perceptually-based Color Assignment}.
\newblock \bibinfo{journal}{\emph{Computer Graphics Forum}}
  \bibinfo{volume}{33}, \bibinfo{number}{7} (\bibinfo{year}{2014}),
  \bibinfo{pages}{309--318}.
\newblock
\href{https://doi.org/10.1111/cgf.12499}{doi:\nolinkurl{10.1111/cgf.12499}}


\bibitem[Kirillov et~al\mbox{.}(2023)]%
        {kirillov2023segment}
\bibfield{author}{\bibinfo{person}{Alexander Kirillov}, \bibinfo{person}{Eric
  Mintun}, \bibinfo{person}{Nikhila Ravi}, \bibinfo{person}{Hanzi Mao},
  \bibinfo{person}{Chloe Rolland}, \bibinfo{person}{Laura Gustafson},
  \bibinfo{person}{Tete Xiao}, \bibinfo{person}{Spencer Whitehead},
  \bibinfo{person}{Alexander~C. Berg}, \bibinfo{person}{Wan-Yen Lo},
  \bibinfo{person}{Piotr Dollar}, {and} \bibinfo{person}{Ross Girshick}.}
  \bibinfo{year}{2023}\natexlab{}.
\newblock \showarticletitle{Segment Anything}. In
  \bibinfo{booktitle}{\emph{Proceedings of the IEEE/CVF International
  Conference on Computer Vision (ICCV)}}. \bibinfo{pages}{4015--4026}.
\newblock
\href{https://doi.org/10.1109/ICCV51070.2023.00371}{doi:\nolinkurl{10.1109/ICCV51070.2023.00371}}


\bibitem[Kobayashi(1981)]%
        {kobayashi1981aim}
\bibfield{author}{\bibinfo{person}{Shigenobu Kobayashi}.}
  \bibinfo{year}{1981}\natexlab{}.
\newblock \showarticletitle{The Aim and Method of the Color Image Scale}.
\newblock \bibinfo{journal}{\emph{Color Research \& Application}}
  \bibinfo{volume}{6}, \bibinfo{number}{2} (\bibinfo{year}{1981}),
  \bibinfo{pages}{93--107}.
\newblock
\href{https://doi.org/10.1002/col.5080060210}{doi:\nolinkurl{10.1002/col.5080060210}}


\bibitem[Kurt and Osueke(2014)]%
        {kurt2014effects}
\bibfield{author}{\bibinfo{person}{Sevinc Kurt} {and}
  \bibinfo{person}{Kelechi~Kingsley Osueke}.} \bibinfo{year}{2014}\natexlab{}.
\newblock \showarticletitle{The Effects of Color on the Moods of College
  Students}.
\newblock \bibinfo{journal}{\emph{Sage Open}} \bibinfo{volume}{4},
  \bibinfo{number}{1} (\bibinfo{year}{2014}).
\newblock
\href{https://doi.org/10.1177/2158244014525423}{doi:\nolinkurl{10.1177/2158244014525423}}


\bibitem[Lee and Hwang(2015)]%
        {lee2015effects}
\bibfield{author}{\bibinfo{person}{Inseok Lee} {and} \bibinfo{person}{Won-Gue
  Hwang}.} \bibinfo{year}{2015}\natexlab{}.
\newblock \showarticletitle{Effects of Personal Experiences on the
  Interpretation of the Meaning of Colours Used in the Displays and Controls in
  Electric Control Panels}.
\newblock \bibinfo{journal}{\emph{Ergonomics}} \bibinfo{volume}{58},
  \bibinfo{number}{12} (\bibinfo{year}{2015}), \bibinfo{pages}{1974--1982}.
\newblock
\href{https://doi.org/10.1080/00140139.2015.1047803}{doi:\nolinkurl{10.1080/00140139.2015.1047803}}


\bibitem[Lee and Lee(2006)]%
        {lee2006development}
\bibfield{author}{\bibinfo{person}{Young-Jin Lee} {and}
  \bibinfo{person}{Joohyeon Lee}.} \bibinfo{year}{2006}\natexlab{}.
\newblock \showarticletitle{The Development of An Emotion Model Based on Colour
  Combinations}.
\newblock \bibinfo{journal}{\emph{International Journal of Consumer Studies}}
  \bibinfo{volume}{30}, \bibinfo{number}{2} (\bibinfo{year}{2006}),
  \bibinfo{pages}{122--136}.
\newblock
\href{https://doi.org/10.1111/j.1470-6431.2005.00457.x}{doi:\nolinkurl{10.1111/j.1470-6431.2005.00457.x}}


\bibitem[Li et~al\mbox{.}(2024)]%
        {li2024sd4match}
\bibfield{author}{\bibinfo{person}{Xinghui Li}, \bibinfo{person}{Jingyi Lu},
  \bibinfo{person}{Kai Han}, {and} \bibinfo{person}{Victor~Adrian Prisacariu}.}
  \bibinfo{year}{2024}\natexlab{}.
\newblock \showarticletitle{SD4Match: Learning to Prompt Stable Diffusion Model
  for Semantic Matching}. In \bibinfo{booktitle}{\emph{Proceedings of the
  IEEE/CVF Conference on Computer Vision and Pattern Recognition (CVPR)}}.
  \bibinfo{pages}{27558--27568}.
\newblock


\bibitem[Lin et~al\mbox{.}(2022)]%
        {lin2022c3}
\bibfield{author}{\bibinfo{person}{Juncong Lin}, \bibinfo{person}{Pintong
  Xiao}, \bibinfo{person}{Yinan Fu}, \bibinfo{person}{Yubin Shi},
  \bibinfo{person}{Hongran Wang}, \bibinfo{person}{Shihui Guo},
  \bibinfo{person}{Ying He}, {and} \bibinfo{person}{Tong-Yee Lee}.}
  \bibinfo{year}{2022}\natexlab{}.
\newblock \showarticletitle{C3 Assignment: Camera Cubemap Color Assignment for
  Creative Interior Design}.
\newblock \bibinfo{journal}{\emph{IEEE Transactions on Visualization and
  Computer Graphics}} \bibinfo{volume}{28}, \bibinfo{number}{8}
  (\bibinfo{year}{2022}), \bibinfo{pages}{2895--2908}.
\newblock
\href{https://doi.org/10.1109/TVCG.2020.3041728}{doi:\nolinkurl{10.1109/TVCG.2020.3041728}}


\bibitem[Lin et~al\mbox{.}(2013)]%
        {lin2013selecting}
\bibfield{author}{\bibinfo{person}{Sharon Lin}, \bibinfo{person}{Julie
  Fortuna}, \bibinfo{person}{Chinmay Kulkarni}, \bibinfo{person}{Maureen
  Stone}, {and} \bibinfo{person}{Jeffrey Heer}.}
  \bibinfo{year}{2013}\natexlab{}.
\newblock \showarticletitle{Selecting Semantically-Resonant Colors for Data
  Visualization}.
\newblock \bibinfo{journal}{\emph{Computer Graphics Forum}}
  \bibinfo{volume}{32}, \bibinfo{number}{3pt4} (\bibinfo{year}{2013}),
  \bibinfo{pages}{401--410}.
\newblock
\showISSN{0167-7055}
\href{https://doi.org/10.1111/cgf.12127}{doi:\nolinkurl{10.1111/cgf.12127}}


\bibitem[Lindner et~al\mbox{.}(2012)]%
        {lindner2012color}
\bibfield{author}{\bibinfo{person}{Albrecht Lindner}, \bibinfo{person}{Nicolas
  Bonnier}, {and} \bibinfo{person}{Sabine S{\"u}sstrunk}.}
  \bibinfo{year}{2012}\natexlab{}.
\newblock \showarticletitle{What is the Color of Chocolate? - Extracting Color
  Values of Semantic Expressions}. In \bibinfo{booktitle}{\emph{Conference on
  Colour in Graphics, Imaging, and Vision}}, Vol.~\bibinfo{volume}{6}. Society
  of Imaging Science and Technology, \bibinfo{pages}{355--361}.
\newblock
\urldef\tempurl%
\url{https://api.semanticscholar.org/CorpusID:2082872}
\showURL{%
\tempurl}


\bibitem[Liu et~al\mbox{.}(2023)]%
        {liu2023grounding}
\bibfield{author}{\bibinfo{person}{Shilong Liu}, \bibinfo{person}{Zhaoyang
  Zeng}, \bibinfo{person}{Tianhe Ren}, \bibinfo{person}{Feng Li},
  \bibinfo{person}{Hao Zhang}, \bibinfo{person}{Jie Yang},
  \bibinfo{person}{Chunyuan Li}, \bibinfo{person}{Jianwei Yang},
  \bibinfo{person}{Hang Su}, \bibinfo{person}{Jun Zhu}, {et~al\mbox{.}}}
  \bibinfo{year}{2023}\natexlab{}.
\newblock \showarticletitle{Grounding DINO: Marrying DINO with Grounded
  Pre-Training for Open-Set Object Detection}.
\newblock \bibinfo{journal}{\emph{arXiv preprint arXiv:2303.05499}}
  (\bibinfo{year}{2023}).
\newblock


\bibitem[Liu and Chilton(2022)]%
        {liu2022design}
\bibfield{author}{\bibinfo{person}{Vivian Liu} {and} \bibinfo{person}{Lydia~B
  Chilton}.} \bibinfo{year}{2022}\natexlab{}.
\newblock \showarticletitle{Design Guidelines for Prompt Engineering
  Text-to-Image Generative Models}. In \bibinfo{booktitle}{\emph{Proceedings of
  the CHI conference on human factors in computing systems}}.
  \bibinfo{pages}{1--23}.
\newblock
\href{https://doi.org/10.1145/3491102.3501825}{doi:\nolinkurl{10.1145/3491102.3501825}}


\bibitem[Lugmayr et~al\mbox{.}(2022)]%
        {lugmayr2022repaint}
\bibfield{author}{\bibinfo{person}{Andreas Lugmayr}, \bibinfo{person}{Martin
  Danelljan}, \bibinfo{person}{Andres Romero}, \bibinfo{person}{Fisher Yu},
  \bibinfo{person}{Radu Timofte}, {and} \bibinfo{person}{Luc Van~Gool}.}
  \bibinfo{year}{2022}\natexlab{}.
\newblock \showarticletitle{RePaint: Inpainting using Denoising Diffusion
  Probabilistic Models}. In \bibinfo{booktitle}{\emph{Proceedings of the
  IEEE/CVF Conference on Computer Vision and Pattern Recognition}}.
  \bibinfo{pages}{11461--11471}.
\newblock
\href{https://doi.org/10.1109/CVPR52688.2022.01117}{doi:\nolinkurl{10.1109/CVPR52688.2022.01117}}


\bibitem[Michel et~al\mbox{.}(2011)]%
        {michel2011quantitative}
\bibfield{author}{\bibinfo{person}{Jean-Baptiste Michel},
  \bibinfo{person}{Yuan~Kui Shen}, \bibinfo{person}{Aviva~Presser Aiden},
  \bibinfo{person}{Adrian Veres}, \bibinfo{person}{Matthew~K. Gray},
  \bibinfo{person}{The Google~Books Team}, \bibinfo{person}{Joseph~P. Pickett},
  \bibinfo{person}{Dale Hoiberg}, \bibinfo{person}{Dan Clancy},
  \bibinfo{person}{Peter Norvig}, \bibinfo{person}{Jon Orwant},
  \bibinfo{person}{Steven Pinker}, \bibinfo{person}{Martin~A. Nowak}, {and}
  \bibinfo{person}{Erez~Lieberman Aiden}.} \bibinfo{year}{2011}\natexlab{}.
\newblock \showarticletitle{Quantitative Analysis of Culture Using Millions of
  Digitized Books}.
\newblock \bibinfo{journal}{\emph{Science}} \bibinfo{volume}{331},
  \bibinfo{number}{6014} (\bibinfo{year}{2011}), \bibinfo{pages}{176--182}.
\newblock
\urldef\tempurl%
\url{http://www.sciencemag.org/content/331/6014/176.full}
\showURL{%
\tempurl}


\bibitem[Mukherjee et~al\mbox{.}(2024)]%
        {mukherjee2024large}
\bibfield{author}{\bibinfo{person}{Kushin Mukherjee},
  \bibinfo{person}{Timothy~T Rogers}, {and} \bibinfo{person}{Karen~B Schloss}.}
  \bibinfo{year}{2024}\natexlab{}.
\newblock \showarticletitle{Large Language Models Estimate Fine-grained Human
  Color-concept Associations}.
\newblock \bibinfo{journal}{\emph{arXiv preprint arXiv:2406.17781}}
  (\bibinfo{year}{2024}).
\newblock


\bibitem[Mukherjee et~al\mbox{.}(2022)]%
        {mukherjee2022context}
\bibfield{author}{\bibinfo{person}{Kushin Mukherjee}, \bibinfo{person}{Brian
  Yin}, \bibinfo{person}{Brianne~E. Sherman}, \bibinfo{person}{Laurent
  Lessard}, {and} \bibinfo{person}{Karen~B. Schloss}.}
  \bibinfo{year}{2022}\natexlab{}.
\newblock \showarticletitle{Context Matters: A Theory of Semantic
  Discriminability for Perceptual Encoding Systems}.
\newblock \bibinfo{journal}{\emph{IEEE Transactions on Visualization and
  Computer Graphics}} \bibinfo{volume}{28}, \bibinfo{number}{1}
  (\bibinfo{year}{2022}), \bibinfo{pages}{697--706}.
\newblock
\href{https://doi.org/10.1109/TVCG.2021.3114780}{doi:\nolinkurl{10.1109/TVCG.2021.3114780}}


\bibitem[M{\"u}ller and Markert(2019)]%
        {muller2019identifying}
\bibfield{author}{\bibinfo{person}{Nicolas~M M{\"u}ller} {and}
  \bibinfo{person}{Karla Markert}.} \bibinfo{year}{2019}\natexlab{}.
\newblock \showarticletitle{Identifying Mislabeled Instances in Classification
  Datasets}. In \bibinfo{booktitle}{\emph{2019 International Joint Conference
  on Neural Networks (IJCNN)}}. \bibinfo{pages}{1--8}.
\newblock
\href{https://doi.org/10.1109/IJCNN.2019.8851920}{doi:\nolinkurl{10.1109/IJCNN.2019.8851920}}


\bibitem[Munroe(2010)]%
        {xkcd2010color}
\bibfield{author}{\bibinfo{person}{Randall Munroe}.}
  \bibinfo{year}{2010}\natexlab{}.
\newblock \bibinfo{title}{Color Survey Results}.
\newblock
\urldef\tempurl%
\url{https://blog.xkcd.com/2010/05/03/color-survey-results/}
\showURL{%
\tempurl}


\bibitem[Naik and Nushi(2023)]%
        {naik2023social}
\bibfield{author}{\bibinfo{person}{Ranjita Naik} {and} \bibinfo{person}{Besmira
  Nushi}.} \bibinfo{year}{2023}\natexlab{}.
\newblock \showarticletitle{Social Biases through the Text-to-Image Generation
  Lens}. In \bibinfo{booktitle}{\emph{Proceedings of the AAAI/ACM Conference on
  AI, Ethics, and Society}} (Montr\'{e}al, QC, Canada)
  \emph{(\bibinfo{series}{AIES '23})}. \bibinfo{publisher}{Association for
  Computing Machinery}, \bibinfo{address}{New York, NY, USA},
  \bibinfo{pages}{786–808}.
\newblock
\showISBNx{9798400702310}
\href{https://doi.org/10.1145/3600211.3604711}{doi:\nolinkurl{10.1145/3600211.3604711}}


\bibitem[Nelson and Stolterman(2014)]%
        {nelson2014design}
\bibfield{author}{\bibinfo{person}{Harold~G Nelson} {and} \bibinfo{person}{Erik
  Stolterman}.} \bibinfo{year}{2014}\natexlab{}.
\newblock \bibinfo{booktitle}{\emph{The Design Way: Intentional Change in An
  Unpredictable World}}.
\newblock \bibinfo{publisher}{MIT press}.
\newblock


\bibitem[Ng and Chan(2018)]%
        {annie2018color}
\bibfield{author}{\bibinfo{person}{Annie~WY Ng} {and} \bibinfo{person}{Alan~HS
  Chan}.} \bibinfo{year}{2018}\natexlab{}.
\newblock \showarticletitle{Color Associations among Designers and
  Non-designers for Common Warning and Operation Concepts}.
\newblock \bibinfo{journal}{\emph{Applied Ergonomics}}  \bibinfo{volume}{70}
  (\bibinfo{year}{2018}), \bibinfo{pages}{18--25}.
\newblock
\showISSN{0003-6870}
\href{https://doi.org/10.1016/j.apergo.2018.02.004}{doi:\nolinkurl{10.1016/j.apergo.2018.02.004}}


\bibitem[Nguyen et~al\mbox{.}(2024)]%
        {nguyen2024dataset}
\bibfield{author}{\bibinfo{person}{Quang Nguyen}, \bibinfo{person}{Truong Vu},
  \bibinfo{person}{Anh Tran}, {and} \bibinfo{person}{Khoi Nguyen}.}
  \bibinfo{year}{2024}\natexlab{}.
\newblock \showarticletitle{Dataset Diffusion: Diffusion-based Synthetic Data
  Generation for Pixel-level Semantic Segmentation}. In
  \bibinfo{booktitle}{\emph{Proceedings of the International Conference on
  Neural Information Processing Systems}}, Vol.~\bibinfo{volume}{36}.
  \bibinfo{publisher}{Curran Associates Inc.}
\newblock


\bibitem[Oppenlaender(2022)]%
        {oppenlaender2022creativity}
\bibfield{author}{\bibinfo{person}{Jonas Oppenlaender}.}
  \bibinfo{year}{2022}\natexlab{}.
\newblock \showarticletitle{The Creativity of Text-to-Image Generation}. In
  \bibinfo{booktitle}{\emph{Proceedings of the 25th International Academic
  Mindtrek Conference}} (Tampere, Finland) \emph{(\bibinfo{series}{Academic
  Mindtrek '22})}. \bibinfo{publisher}{Association for Computing Machinery},
  \bibinfo{address}{New York, NY, USA}, \bibinfo{pages}{192–202}.
\newblock
\showISBNx{9781450399555}
\href{https://doi.org/10.1145/3569219.3569352}{doi:\nolinkurl{10.1145/3569219.3569352}}


\bibitem[Orgad et~al\mbox{.}(2023)]%
        {orgad2023editing}
\bibfield{author}{\bibinfo{person}{Hadas Orgad}, \bibinfo{person}{Bahjat
  Kawar}, {and} \bibinfo{person}{Yonatan Belinkov}.}
  \bibinfo{year}{2023}\natexlab{}.
\newblock \showarticletitle{Editing Implicit Assumptions in Text-to-Image
  Diffusion Models}. In \bibinfo{booktitle}{\emph{2023 IEEE/CVF International
  Conference on Computer Vision (ICCV)}}. \bibinfo{pages}{7030--7038}.
\newblock
\href{https://doi.org/10.1109/ICCV51070.2023.00649}{doi:\nolinkurl{10.1109/ICCV51070.2023.00649}}


\bibitem[Osgood(1957)]%
        {osgood1957measurement}
\bibfield{author}{\bibinfo{person}{Charles~E Osgood}.}
  \bibinfo{year}{1957}\natexlab{}.
\newblock \bibinfo{booktitle}{\emph{The Measurement of Meaning}}.
\newblock \bibinfo{publisher}{University of Illinois press}.
\newblock


\bibitem[Ou et~al\mbox{.}(2004)]%
        {ou2004study}
\bibfield{author}{\bibinfo{person}{Li-Chen Ou}, \bibinfo{person}{M.~Ronnier
  Luo}, \bibinfo{person}{Andrée Woodcock}, {and} \bibinfo{person}{Angela
  Wright}.} \bibinfo{year}{2004}\natexlab{}.
\newblock \showarticletitle{A Study of Colour Emotion and Colour Preference.
  Part II: Colour Emotions for Two-colour Combinations}.
\newblock \bibinfo{journal}{\emph{Color Research \& Application}}
  \bibinfo{volume}{29}, \bibinfo{number}{4} (\bibinfo{year}{2004}),
  \bibinfo{pages}{292--298}.
\newblock
\href{https://doi.org/10.1002/col.20024}{doi:\nolinkurl{10.1002/col.20024}}


\bibitem[Palmer and Schloss(2010)]%
        {palmer2010ecological}
\bibfield{author}{\bibinfo{person}{Stephen~E. Palmer} {and}
  \bibinfo{person}{Karen~B. Schloss}.} \bibinfo{year}{2010}\natexlab{}.
\newblock \showarticletitle{An Ecological Valence Theory of Human Color
  Preference}.
\newblock \bibinfo{journal}{\emph{Proceedings of the National Academy of
  Sciences}} \bibinfo{volume}{107}, \bibinfo{number}{19}
  (\bibinfo{year}{2010}), \bibinfo{pages}{8877--8882}.
\newblock
\href{https://doi.org/10.1073/pnas.0906172107}{doi:\nolinkurl{10.1073/pnas.0906172107}}


\bibitem[Parsons(2022)]%
        {dalleprompt2022guy}
\bibfield{author}{\bibinfo{person}{Guy Parsons}.}
  \bibinfo{year}{2022}\natexlab{}.
\newblock \bibinfo{title}{The DALL·E 2 Prompt Book}.
\newblock
\urldef\tempurl%
\url{https://dallery.gallery/the-dalle-2-prompt-book/}
\showURL{%
\tempurl}


\bibitem[Prerak(2024)]%
        {prerak2024addressing}
\bibfield{author}{\bibinfo{person}{Shah Prerak}.}
  \bibinfo{year}{2024}\natexlab{}.
\newblock \showarticletitle{Addressing Bias in Text-to-Image Generation: A
  Review of Mitigation Methods}. In \bibinfo{booktitle}{\emph{Proceedings of
  the International Conference on Smart Technologies and Systems for Next
  Generation Computing (ICSTSN)}}. \bibinfo{pages}{1--6}.
\newblock
\href{https://doi.org/10.1109/ICSTSN61422.2024.10671230}{doi:\nolinkurl{10.1109/ICSTSN61422.2024.10671230}}


\bibitem[Qiu et~al\mbox{.}(2022)]%
        {qiu2022intelligent}
\bibfield{author}{\bibinfo{person}{Qianru Qiu}, \bibinfo{person}{Mayu Otani},
  {and} \bibinfo{person}{Yuki Iwazaki}.} \bibinfo{year}{2022}\natexlab{}.
\newblock \showarticletitle{An Intelligent Color Recommendation Tool for
  Landing Page Design}. In \bibinfo{booktitle}{\emph{Companion Proceedings of
  the International Conference on Intelligent User Interfaces}}.
  \bibinfo{publisher}{Association for Computing Machinery},
  \bibinfo{pages}{26–29}.
\newblock
\href{https://doi.org/10.1145/3490100.3516450}{doi:\nolinkurl{10.1145/3490100.3516450}}


\bibitem[Rathore et~al\mbox{.}(2020)]%
        {rathore2020estimating}
\bibfield{author}{\bibinfo{person}{Ragini Rathore}, \bibinfo{person}{Zachary
  Leggon}, \bibinfo{person}{Laurent Lessard}, {and} \bibinfo{person}{Karen~B.
  Schloss}.} \bibinfo{year}{2020}\natexlab{}.
\newblock \showarticletitle{Estimating Color-Concept Associations from Image
  Statistics}.
\newblock \bibinfo{journal}{\emph{IEEE Transactions on Visualization and
  Computer Graphics}} \bibinfo{volume}{26}, \bibinfo{number}{1}
  (\bibinfo{year}{2020}), \bibinfo{pages}{1226--1235}.
\newblock
\href{https://doi.org/10.1109/TVCG.2019.2934536}{doi:\nolinkurl{10.1109/TVCG.2019.2934536}}


\bibitem[Ren et~al\mbox{.}(2024)]%
        {ren2024grounded}
\bibfield{author}{\bibinfo{person}{Tianhe Ren}, \bibinfo{person}{Shilong Liu},
  \bibinfo{person}{Ailing Zeng}, \bibinfo{person}{Jing Lin},
  \bibinfo{person}{Kunchang Li}, \bibinfo{person}{He Cao},
  \bibinfo{person}{Jiayu Chen}, \bibinfo{person}{Xinyu Huang},
  \bibinfo{person}{Yukang Chen}, \bibinfo{person}{Feng Yan},
  \bibinfo{person}{Zhaoyang Zeng}, \bibinfo{person}{Hao Zhang},
  \bibinfo{person}{Feng Li}, \bibinfo{person}{Jie Yang},
  \bibinfo{person}{Hongyang Li}, \bibinfo{person}{Qing Jiang}, {and}
  \bibinfo{person}{Lei Zhang}.} \bibinfo{year}{2024}\natexlab{}.
\newblock \showarticletitle{Grounded SAM: Assembling Open-World Models for
  Diverse Visual Tasks}.
\newblock \bibinfo{journal}{\emph{arXiv preprint arXiv:2401.14159}}
  (\bibinfo{year}{2024}).
\newblock


\bibitem[Rombach et~al\mbox{.}(2022)]%
        {rombach2022high}
\bibfield{author}{\bibinfo{person}{Robin Rombach}, \bibinfo{person}{Andreas
  Blattmann}, \bibinfo{person}{Dominik Lorenz}, \bibinfo{person}{Patrick
  Esser}, {and} \bibinfo{person}{Bj{\"o}rn Ommer}.}
  \bibinfo{year}{2022}\natexlab{}.
\newblock \showarticletitle{High-Resolution Image Synthesis With Latent
  Diffusion Models}. In \bibinfo{booktitle}{\emph{Proceedings of the IEEE/CVF
  conference on computer vision and pattern recognition}}.
  \bibinfo{pages}{10684--10695}.
\newblock


\bibitem[Ruiz et~al\mbox{.}(2023)]%
        {ruiz2023dreambooth}
\bibfield{author}{\bibinfo{person}{Nataniel Ruiz}, \bibinfo{person}{Yuanzhen
  Li}, \bibinfo{person}{Varun Jampani}, \bibinfo{person}{Yael Pritch},
  \bibinfo{person}{Michael Rubinstein}, {and} \bibinfo{person}{Kfir Aberman}.}
  \bibinfo{year}{2023}\natexlab{}.
\newblock \showarticletitle{DreamBooth: Fine Tuning Text-to-Image Diffusion
  Models for Subject-Driven Generation}. In \bibinfo{booktitle}{\emph{2023
  IEEE/CVF Conference on Computer Vision and Pattern Recognition (CVPR)}}.
  \bibinfo{pages}{22500--22510}.
\newblock
\href{https://doi.org/10.1109/CVPR52729.2023.02155}{doi:\nolinkurl{10.1109/CVPR52729.2023.02155}}


\bibitem[Saharia et~al\mbox{.}(2022)]%
        {saharia2022photorealistic}
\bibfield{author}{\bibinfo{person}{Chitwan Saharia}, \bibinfo{person}{William
  Chan}, \bibinfo{person}{Saurabh Saxena}, \bibinfo{person}{Lala Lit},
  \bibinfo{person}{Jay Whang}, \bibinfo{person}{Emily Denton},
  \bibinfo{person}{Seyed Kamyar~Seyed Ghasemipour},
  \bibinfo{person}{Burcu~Karagol Ayan}, \bibinfo{person}{S.~Sara Mahdavi},
  \bibinfo{person}{Raphael Gontijo-Lopes}, \bibinfo{person}{Tim Salimans},
  \bibinfo{person}{Jonathan Ho}, \bibinfo{person}{David~J Fleet}, {and}
  \bibinfo{person}{Mohammad Norouzi}.} \bibinfo{year}{2022}\natexlab{}.
\newblock \showarticletitle{Photorealistic Text-to-Image Diffusion Models with
  Deep Language Understanding}. In \bibinfo{booktitle}{\emph{Proceedings of the
  Advances in Neural Information Processing Systems}}.
  \bibinfo{pages}{36479--36494}.
\newblock
\urldef\tempurl%
\url{https://proceedings.neurips.cc/paper_files/paper/2022/file/ec795aeadae0b7d230fa35cbaf04c041-Paper-Conference.pdf}
\showURL{%
\tempurl}


\bibitem[Samuelson(2023)]%
        {samuelson2023generative}
\bibfield{author}{\bibinfo{person}{Pamela Samuelson}.}
  \bibinfo{year}{2023}\natexlab{}.
\newblock \showarticletitle{Generative AI meets copyright}.
\newblock \bibinfo{journal}{\emph{Science}} \bibinfo{volume}{381},
  \bibinfo{number}{6654} (\bibinfo{year}{2023}), \bibinfo{pages}{158--161}.
\newblock
\href{https://doi.org/10.1126/science.adi0656}{doi:\nolinkurl{10.1126/science.adi0656}}
\showeprint{https://www.science.org/doi/pdf/10.1126/science.adi0656}


\bibitem[Sandnes(2021)]%
        {sandnes2021inverse}
\bibfield{author}{\bibinfo{person}{Frode~Eika Sandnes}.}
  \bibinfo{year}{2021}\natexlab{}.
\newblock \showarticletitle{Inverse Color Contrast Checker: Automatically
  Suggesting Color Adjustments that meet Contrast Requirements on the Web}. In
  \bibinfo{booktitle}{\emph{Proceedings of the International ACM SIGACCESS
  Conference on Computers and Accessibility}} (Virtual Event, USA)
  \emph{(\bibinfo{series}{ASSETS '21})}. \bibinfo{publisher}{Association for
  Computing Machinery}, \bibinfo{address}{New York, NY, USA}, Article
  \bibinfo{articleno}{72}, \bibinfo{numpages}{4}~pages.
\newblock
\showISBNx{9781450383066}
\href{https://doi.org/10.1145/3441852.3476529}{doi:\nolinkurl{10.1145/3441852.3476529}}


\bibitem[Sar{\i}y{\i}ld{\i}z et~al\mbox{.}(2023)]%
        {sariyildiz2023fake}
\bibfield{author}{\bibinfo{person}{Mert~B{\"u}lent Sar{\i}y{\i}ld{\i}z},
  \bibinfo{person}{Karteek Alahari}, \bibinfo{person}{Diane Larlus}, {and}
  \bibinfo{person}{Yannis Kalantidis}.} \bibinfo{year}{2023}\natexlab{}.
\newblock \showarticletitle{Fake it Till You Make it: Learning Transferable
  Representations from Synthetic ImageNet Clones}. In
  \bibinfo{booktitle}{\emph{Proceedings of the IEEE/CVF Conference on Computer
  Vision and Pattern Recognition}}. \bibinfo{pages}{8011--8021}.
\newblock
\href{https://doi.org/10.1109/CVPR52729.2023.00774}{doi:\nolinkurl{10.1109/CVPR52729.2023.00774}}


\bibitem[Schloss et~al\mbox{.}(2018)]%
        {schloss2018color}
\bibfield{author}{\bibinfo{person}{Karen~B Schloss}, \bibinfo{person}{Laurent
  Lessard}, \bibinfo{person}{Charlotte~S Walmsley}, {and}
  \bibinfo{person}{Kathleen Foley}.} \bibinfo{year}{2018}\natexlab{}.
\newblock \showarticletitle{Color Inference in Visual Communication: The
  Meaning of Colors in Recycling}.
\newblock \bibinfo{journal}{\emph{Cognitive Research: Principles and
  Implications}}  \bibinfo{volume}{3} (\bibinfo{year}{2018}),
  \bibinfo{pages}{1--17}.
\newblock
\href{https://doi.org/10.1186/s41235-018-0090-y}{doi:\nolinkurl{10.1186/s41235-018-0090-y}}


\bibitem[Seo-young Lee and Hoffman(2024)]%
        {lee2024when}
\bibfield{author}{\bibinfo{person}{Matthew~Law Seo-young Lee} {and}
  \bibinfo{person}{Guy Hoffman}.} \bibinfo{year}{2024}\natexlab{}.
\newblock \showarticletitle{When and How to Use AI in the Design Process?
  Implications for Human-AI Design Collaboration}.
\newblock \bibinfo{journal}{\emph{International Journal of Human–Computer
  Interaction}} \bibinfo{volume}{0}, \bibinfo{number}{0}
  (\bibinfo{year}{2024}), \bibinfo{pages}{1--16}.
\newblock
\href{https://doi.org/10.1080/10447318.2024.2353451}{doi:\nolinkurl{10.1080/10447318.2024.2353451}}


\bibitem[Setlur and Stone(2016)]%
        {setlur2016linguistic}
\bibfield{author}{\bibinfo{person}{V. Setlur} {and} \bibinfo{person}{M.~C.
  Stone}.} \bibinfo{year}{2016}\natexlab{}.
\newblock \showarticletitle{A Linguistic Approach to Categorical Color
  Assignment for Data Visualization}.
\newblock \bibinfo{journal}{\emph{IEEE Transactions on Visualization and
  Computer Graphics}} \bibinfo{volume}{22}, \bibinfo{number}{1}
  (\bibinfo{year}{2016}), \bibinfo{pages}{698--707}.
\newblock
\showISSN{1077-2626}
\href{https://doi.org/10.1109/TVCG.2015.2467471}{doi:\nolinkurl{10.1109/TVCG.2015.2467471}}


\bibitem[Shen et~al\mbox{.}(2000)]%
        {shen2000color}
\bibfield{author}{\bibinfo{person}{Yu-Chuan Shen}, \bibinfo{person}{Wu-Hsiung
  Yuan}, \bibinfo{person}{Wen-Hsing Hsu}, {and} \bibinfo{person}{Yung-Sheng
  Chen}.} \bibinfo{year}{2000}\natexlab{}.
\newblock \showarticletitle{Color Selection in The Consideration of Color
  Harmony for Interior Design}.
\newblock \bibinfo{journal}{\emph{Color Research and Application}}
  \bibinfo{volume}{25}, \bibinfo{number}{1} (\bibinfo{year}{2000}),
  \bibinfo{pages}{20--31}.
\newblock
\href{https://doi.org/10.1002/(SICI)1520-6378(200002)25:1<20::AID-COL4>3.0.CO;2-5}{doi:\nolinkurl{10.1002/(SICI)1520-6378(200002)25:1<20::AID-COL4>3.0.CO;2-5}}


\bibitem[Shi et~al\mbox{.}(2023)]%
        {shi2023stijl}
\bibfield{author}{\bibinfo{person}{Xinyu Shi}, \bibinfo{person}{Ziqi Zhou},
  \bibinfo{person}{Jing~Wen Zhang}, \bibinfo{person}{Ali Neshati},
  \bibinfo{person}{Anjul~Kumar Tyagi}, \bibinfo{person}{Ryan Rossi},
  \bibinfo{person}{Shunan Guo}, \bibinfo{person}{Fan Du}, {and}
  \bibinfo{person}{Jian Zhao}.} \bibinfo{year}{2023}\natexlab{}.
\newblock \showarticletitle{De-Stijl: Facilitating Graphics Design with
  Interactive 2D Color Palette Recommendation}. In
  \bibinfo{booktitle}{\emph{Proceedings of the CHI Conference on Human Factors
  in Computing Systems}}. \bibinfo{publisher}{Association for Computing
  Machinery}, Article \bibinfo{articleno}{122}, \bibinfo{numpages}{19}~pages.
\newblock
\showISBNx{9781450394215}
\href{https://doi.org/10.1145/3544548.3581070}{doi:\nolinkurl{10.1145/3544548.3581070}}


\bibitem[Shi et~al\mbox{.}(2022)]%
        {shi2022colorcook}
\bibfield{author}{\bibinfo{person}{Yang Shi}, \bibinfo{person}{Siji Chen},
  \bibinfo{person}{Pei Liu}, \bibinfo{person}{Jiang Long}, {and}
  \bibinfo{person}{Nan Cao}.} \bibinfo{year}{2022}\natexlab{}.
\newblock \showarticletitle{ColorCook: Augmenting Color Design for Dashboarding
  with Domain-Associated Palettes}.
\newblock \bibinfo{journal}{\emph{Proc. ACM Hum.-Comput. Interact.}}
  \bibinfo{volume}{6}, \bibinfo{number}{CSCW2}, Article
  \bibinfo{articleno}{433} (\bibinfo{date}{nov} \bibinfo{year}{2022}),
  \bibinfo{numpages}{25}~pages.
\newblock
\href{https://doi.org/10.1145/3555534}{doi:\nolinkurl{10.1145/3555534}}


\bibitem[Shi et~al\mbox{.}(2024)]%
        {shi2024personalizing}
\bibfield{author}{\bibinfo{person}{Yang Shi}, \bibinfo{person}{Yechun Peng},
  \bibinfo{person}{Shengqi Dang}, \bibinfo{person}{Nanxuan Zhao}, {and}
  \bibinfo{person}{Nan Cao}.} \bibinfo{year}{2024}\natexlab{}.
\newblock \showarticletitle{Personalizing Products with Stylized Head Portraits
  for Self-Expression}. In \bibinfo{booktitle}{\emph{Proceedings of the CHI
  Conference on Human Factors in Computing Systems}}. \bibinfo{pages}{1--18}.
\newblock
\href{https://doi.org/10.1145/3613904.3642391}{doi:\nolinkurl{10.1145/3613904.3642391}}


\bibitem[Siglidis et~al\mbox{.}(2024)]%
        {ioannis2024diffusion}
\bibfield{author}{\bibinfo{person}{Ioannis Siglidis},
  \bibinfo{person}{Aleksander Holynski}, \bibinfo{person}{Alexei~A Efros},
  \bibinfo{person}{Mathieu Aubry}, {and} \bibinfo{person}{Shiry Ginosar}.}
  \bibinfo{year}{2024}\natexlab{}.
\newblock \showarticletitle{Diffusion Models as Data Mining Tools}.
\newblock \bibinfo{journal}{\emph{arXiv preprint arXiv:2408.02752}}
  (\bibinfo{year}{2024}).
\newblock


\bibitem[Szafir(2018)]%
        {szafir2017modeling}
\bibfield{author}{\bibinfo{person}{Danielle~Albers Szafir}.}
  \bibinfo{year}{2018}\natexlab{}.
\newblock \showarticletitle{Modeling Color Difference for Visualization
  Design}.
\newblock \bibinfo{journal}{\emph{IEEE Transactions on Visualization and
  Computer Graphics}} \bibinfo{volume}{24}, \bibinfo{number}{1}
  (\bibinfo{year}{2018}), \bibinfo{pages}{392--401}.
\newblock
\href{https://doi.org/10.1109/TVCG.2017.2744359}{doi:\nolinkurl{10.1109/TVCG.2017.2744359}}


\bibitem[Tham et~al\mbox{.}(2020)]%
        {tham2020systematic}
\bibfield{author}{\bibinfo{person}{Diana Su~Yun Tham}, \bibinfo{person}{Paul~T
  Sowden}, \bibinfo{person}{Alexandra Grandison}, \bibinfo{person}{Anna
  Franklin}, \bibinfo{person}{Anna Kai~Win Lee}, \bibinfo{person}{Michelle Ng},
  \bibinfo{person}{Juhyun Park}, \bibinfo{person}{Weiguo Pang}, {and}
  \bibinfo{person}{Jingwen Zhao}.} \bibinfo{year}{2020}\natexlab{}.
\newblock \showarticletitle{A Systematic Investigation of Conceptual Color
  Associations}.
\newblock \bibinfo{journal}{\emph{Journal of Experimental Psychology: General}}
  \bibinfo{volume}{149}, \bibinfo{number}{7} (\bibinfo{year}{2020}),
  \bibinfo{pages}{1311}.
\newblock
\href{https://doi.org/10.1037/xge0000703}{doi:\nolinkurl{10.1037/xge0000703}}


\bibitem[Verheijden and Funk(2023)]%
        {verheijden2023collaborative}
\bibfield{author}{\bibinfo{person}{Mathias~Peter Verheijden} {and}
  \bibinfo{person}{Mathias Funk}.} \bibinfo{year}{2023}\natexlab{}.
\newblock \showarticletitle{Collaborative Diffusion: Boosting Designerly
  Co-Creation with Generative AI}. In \bibinfo{booktitle}{\emph{Extended
  abstracts of the 2023 CHI conference on human factors in computing systems}}.
  \bibinfo{pages}{1--8}.
\newblock
\href{https://doi.org/10.1145/3544549.3585680}{doi:\nolinkurl{10.1145/3544549.3585680}}


\bibitem[Wan et~al\mbox{.}(2024)]%
        {wan2024survey}
\bibfield{author}{\bibinfo{person}{Yixin Wan}, \bibinfo{person}{Arjun
  Subramonian}, \bibinfo{person}{Anaelia Ovalle}, \bibinfo{person}{Zongyu Lin},
  \bibinfo{person}{Ashima Suvarna}, \bibinfo{person}{Christina Chance},
  \bibinfo{person}{Hritik Bansal}, \bibinfo{person}{Rebecca Pattichis}, {and}
  \bibinfo{person}{Kai-Wei Chang}.} \bibinfo{year}{2024}\natexlab{}.
\newblock \showarticletitle{Survey of Bias In Text-to-Image Generation:
  Definition, Evaluation, and Mitigation}.
\newblock \bibinfo{journal}{\emph{arXiv preprint arXiv:2404.01030}}
  (\bibinfo{year}{2024}).
\newblock


\bibitem[WIERZBICKA(1990)]%
        {wierzicka1990meaning}
\bibfield{author}{\bibinfo{person}{ANNA WIERZBICKA}.}
  \bibinfo{year}{1990}\natexlab{}.
\newblock \showarticletitle{The meaning of color terms: semantics, culture, and
  cognition}.
\newblock \bibinfo{journal}{\emph{Cognitive Linguistics}} \bibinfo{volume}{1},
  \bibinfo{number}{1} (\bibinfo{year}{1990}), \bibinfo{pages}{99--150}.
\newblock
\href{https://doi.org/doi:10.1515/cogl.1990.1.1.99}{doi:\nolinkurl{doi:10.1515/cogl.1990.1.1.99}}


\bibitem[Witteveen and Andrews(2022)]%
        {witteveen2022investigating}
\bibfield{author}{\bibinfo{person}{Sam Witteveen} {and} \bibinfo{person}{Martin
  Andrews}.} \bibinfo{year}{2022}\natexlab{}.
\newblock \showarticletitle{Investigating Prompt Engineering in Diffusion
  Models}.
\newblock \bibinfo{journal}{\emph{arXiv preprint arXiv:2211.15462}}
  (\bibinfo{year}{2022}).
\newblock


\bibitem[Wright and Rainwater(1962)]%
        {wright1962meanings}
\bibfield{author}{\bibinfo{person}{Benjamin Wright} {and} \bibinfo{person}{Lee
  Rainwater}.} \bibinfo{year}{1962}\natexlab{}.
\newblock \showarticletitle{The Meanings of Color}.
\newblock \bibinfo{journal}{\emph{The Journal of General Psychology}}
  \bibinfo{volume}{67}, \bibinfo{number}{1} (\bibinfo{year}{1962}),
  \bibinfo{pages}{89--99}.
\newblock


\bibitem[Xiao et~al\mbox{.}(2024)]%
        {xiao2024typedance}
\bibfield{author}{\bibinfo{person}{Shishi Xiao}, \bibinfo{person}{Liangwei
  Wang}, \bibinfo{person}{Xiaojuan Ma}, {and} \bibinfo{person}{Wei Zeng}.}
  \bibinfo{year}{2024}\natexlab{}.
\newblock \showarticletitle{TypeDance: Creating semantic typographic logos from
  image through personalized generation}. In
  \bibinfo{booktitle}{\emph{Proceedings of the CHI Conference on Human Factors
  in Computing Systems}}. \bibinfo{pages}{1--18}.
\newblock
\href{https://doi.org/10.1145/3613904.3642185}{doi:\nolinkurl{10.1145/3613904.3642185}}


\bibitem[Zhang et~al\mbox{.}(2023)]%
        {zhang2023adding}
\bibfield{author}{\bibinfo{person}{Lvmin Zhang}, \bibinfo{person}{Anyi Rao},
  {and} \bibinfo{person}{Maneesh Agrawala}.} \bibinfo{year}{2023}\natexlab{}.
\newblock \showarticletitle{Adding Conditional Control to Text-to-Image
  Diffusion Models}. In \bibinfo{booktitle}{\emph{Proceedings of the IEEE/CVF
  International Conference on Computer Vision}}. \bibinfo{pages}{3836--3847}.
\newblock


\end{thebibliography}


\newpage
\appendix

\section{Sensitivity of the Diffusion Model}
In our study, we observed that the T2I model exhibits varying sensitivity to different types of prompts.
Specifically, the model is more responsive to prompts with visual descriptions compared to abstract or non-visual concepts.
To address this, we enhanced the prompts by incorporating related visual descriptions, which helped the model better capture the intended context.
Instead of only use \q{quiet}, we added \q{evoking feelings of silence and lonely} to provide a more concrete visual description.

\begin{figure*}[b]
    \centering
    \includegraphics[width=0.9\linewidth]{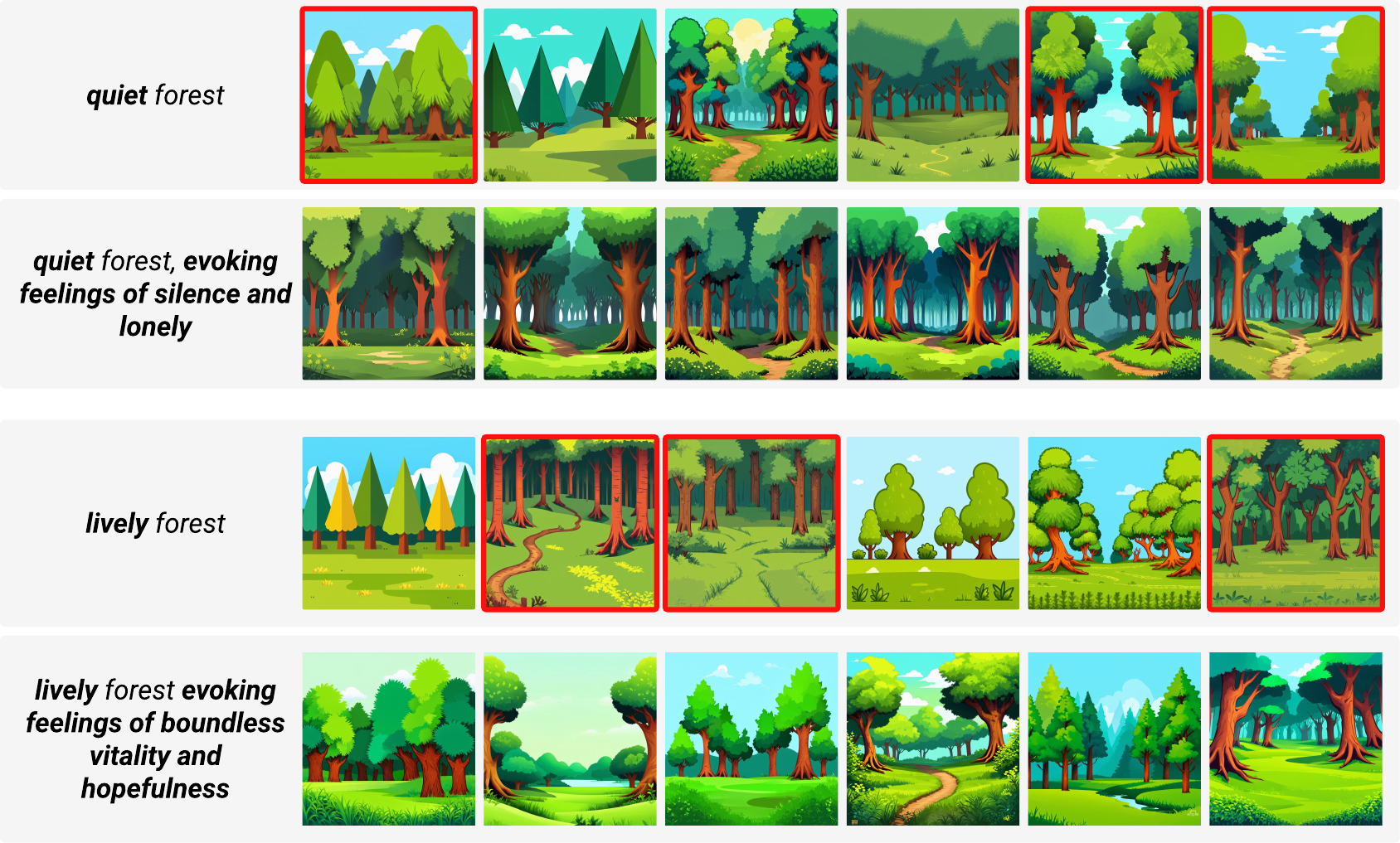}
    \caption{Sensitivity of the T2I model to different types of prompts, comparing the original prompts with enhanced prompts. The red boxes highlight images that do not align with the intended concept.}
    \Description{Sensitivity of the T2I model to different types of prompts, comparing the original prompts with enhanced prompts. The red boxes highlight images that do not align with the intended concept.}
    \label{fig:appendix_figure_1}
\end{figure*}

Figure~\ref{fig:appendix_figure_1} demonstrates this effect. The top row, prompted with "quiet forest," includes images that, despite the prompt, exhibit open and sunlit scenes (highlighted in red), which contradict the expected mood.
The second row, with the enhanced prompt, shows a more consistent adherence to the quiet and solitary atmosphere.
Similarly, the bottom two rows compare "lively forest" with its enhanced version; the latter produces images with a more vibrant and hopeful tone, as intended, with the red boxes indicating images that still convey a sense of depth and solitude rather than liveliness.

\end{document}